\RequirePackage[orthodox,l2tabu]{nag}

\documentclass[reprint,showpacs,superscriptaddress,aps,floatfix,prb]{revtex4-2}
\usepackage[T1]{fontenc}        
\usepackage[utf8]{inputenc}
\usepackage[english]{babel}
\usepackage{dsfont}

\usepackage[left=2.7cm,
            right=2.7cm,
            top=2.7cm,
            bottom=3.2cm
            ]{geometry}        

\usepackage[protrusion,
            expansion
            ]{microtype} 

\usepackage{booktabs}    

\usepackage{enumitem}

\usepackage{appendix}
\usepackage{fancyhdr}

\usepackage{braket}
\usepackage{xspace}
\usepackage{mathdots}
\usepackage{stackrel}
\usepackage{xcolor}
\usepackage{mathtools}
\usepackage{graphicx}
\usepackage{amsmath,          
            amssymb,          
            amsthm}          
\usepackage{xfrac}
\usepackage{comment}

\usepackage{mathrsfs}
\usepackage{bbm}      
\usepackage{tikz}

\definecolor{mycolor1}{HTML}{1F7A8C}
\definecolor{mycolor2}{HTML}{2D3362}
\definecolor{mycolor3}{HTML}{BF1363}
\definecolor{mycolor4}{HTML}{F39237}
\definecolor{mycolor5}{HTML}{2D8B6A}

\usepackage[colorlinks,
            linkcolor=black,
            citecolor=black,
            urlcolor=mycolor3,
            bookmarks,
            bookmarksnumbered
            ]{hyperref}    

\usepackage{cleveref}
\crefname{figure}{Fig.}{Figs.}
\crefname{equation}{Eq.}{Eqs.}
\crefname{section}{Sec.}{Secs.}

\allowdisplaybreaks

\theoremstyle{definition}

\theoremstyle{remark}

\newcommand{\CC}{\mathbb{C}}

\newcommand{\ii}{\mathrm{i}}

\newcommand{\hilb}{\mathcal{H}}
\newcommand{\transf}{\mathcal{T}}

\newcommand{\avgE}{\overline{E}}

\newcommand{\abs}[1]{\left\vert#1\right\vert}

\DeclareMathOperator{\arccosh}{arccosh}

\DeclareMathOperator{\Tr}{Tr}   

\newcommand{\fed}[1]{{\color{black}  #1}}
\newcommand{\redd}[1]{{\color{black}  #1}}

\begin{document}

\title{Eigenstate Thermalization Hypothesis (ETH)  for off-diagonal matrix elements in integrable spin chains}
\author{Federico Rottoli}
\author{Vincenzo Alba}
\affiliation{Dipartimento di Fisica dell’Universit\`a di Pisa and INFN Sezione di Pisa, Largo B. Pontecorvo 3, I-56127 Pisa, Italy.}

\begin{abstract}
We investigate off-diagonal matrix elements of local operators in integrable spin chains, 
focusing on the isotropic spin-$1/2$ Heisenberg chain ($XXX$ chain). We employ state-of-the-art Algebraic Bethe Ansatz results, which allow us to efficiently compute matrix elements of operators with support up to two sites between generic energy eigenstates. We consider both matrix elements between eigenstates that are in the same thermodynamic macrostate, as well as eigenstates that belong to  different macrostates. In the former case, focusing on thermal states we numerically show that matrix elements  
are compatible with the exponential decay as $\exp(-L |{M}^{\scriptscriptstyle{\mathcal{O}}}_{ij}|)$. The probability distribution functions of ${M}_{ij}^{\scriptscriptstyle{\mathcal{O}}}$ depend on the observable and on the 
macrostate, and are well described by Gumbel distributions. On the other hand, matrix elements between eigenstates in different macrostates decay faster as $\exp(-|{M'}_{ij}^{\scriptscriptstyle{\mathcal{O}}}|L^2)$, with ${M'}_{ij}^{\scriptscriptstyle \mathcal{O}}$, again, compatible with a Gumbel distribution. 
\end{abstract}

\maketitle

\section{Introduction}\label{sec:intro}

A fundamental problem in the study of out-of-equilibrium quantum many-body physics is to understand 
how the unitary evolution of isolated systems leads to local equilibration and thermalization. 
Our current understanding of thermalization is based on the celebrated Eigenstate Thermalization Hypothesis~\cite{Deutsch:1991msp,Srednicki:1994mfb,Srednicki:1999bhx,Dalessio2016from} (\emph{ETH}). 
Let $H$ be a generic, i.e., chaotic, many-body Hamiltonian and let $\ket{E_i}$ be its energy eigenstates $H\ket{E_i} = E_i\ket{E_i}$.
According to \emph{ETH}, the matrix elements of a generic local operator ${\mathcal{O}}$ between energy eigenstates take the form~\cite{Deutsch:1991msp,Srednicki:1994mfb,Srednicki:1999bhx,Dalessio2016from}
\begin{equation}\label{eq:ETH}
    \bra{E_i} {\mathcal{O}} \ket{E_j} = O(\avgE)\, \delta_{ij} + f_{\mathcal{O}}(\omega, \avgE)\, e^{-S(\avgE)/2}R_{ij}\,,
\end{equation}
where $\avgE = \frac{1}{2}(E_i + E_j)$ is the average energy of the eigenstates, $\omega = \abs{E_i - E_j}$, $O(\avgE)$ is a smooth function of the energy $\avgE$. In~\eqref{eq:ETH}, $S(\avgE)$ is the thermodynamic  entropy at $\bar E$, $f_{\mathcal{O}}(\omega, \avgE)$ is a smooth function of its arguments, and $R_{ij}$ is a random variable with mean $\overline{R_{ij}} = 0$ and variance $\overline{R_{ij}^2}= 1$. 
Eq.~\eqref{eq:ETH} implies that in the thermodynamic limit, since the entropy $S$ is extensive in the system size $L$, diagonal matrix elements converge  to $O(\avgE)$, i.e., they become smooth functions of the energy \emph{only}. Moreover,  the variance of diagonal matrix elements decays exponentially with $L$. Analogously, off-diagonal matrix elements are exponentially suppressed with increasing $L$.  An intriguing recent result is that 
the statistics of the off-diagonal matrix elements can be understood in terms of Free Probability Theory~\cite{pappalardi2022eigenstate,bernard2023exact}. Crucially, Eq.~\eqref{eq:ETH} implies local thermalization after quantum quenches in generic, i.e., nonintegrable systems, at least for typical initial states~\cite{Dalessio2016from}. 
The \emph{ETH} scaling~\eqref{eq:ETH}  has been checked numerically for both diagonal and off-diagonal matrix elements in nonintegrable systems by employing exact diagonalization methods~\cite{rigol2008thermalization,steinigeweg2013eigenstate,beugeling2015off,chandran2016the,mondaini2017eigenstate,nation2018off,yoshizawa2018numerical,rigol2010quantum}. 
In integrable models, the presence of an extensive number of conserved charges leads to violations of~\eqref{eq:ETH}, which have been investigated numerically~\cite{biroli2010effect,khatami2013fluctuations,leblond2019entanglement,brenes2020low,leblond2020eigenstate,mierzejewski2020quantitative,zhang2022statistical,patil2025eigenstate}. Refs.~\cite{ikeda2013finite,Alba:2009th} investigated the \emph{ETH} scaling of diagonal matrix element in \emph{interacting} integrable systems by exploiting Algebraic Bethe Ansatz (\emph{ABA}) results~\cite{ikeda2013finite,AlbaETHdiagonal} for correlation functions, which allow to reach much larger 
system sizes compared with exact diagonalization. 
Precisely, Ref.~\cite{AlbaETHdiagonal} showed that the variance of \emph{diagonal} matrix elements decays as 
$L^{-1}$, in contrast with~\eqref{eq:ETH}. 

Very recently, Ref.~\cite{Essler:2023nhu} investigated off-diagonal matrix elements of both local and non-local operators in the Lieb-Liniger model, which describes a one-dimensional system of interacting bosons in the continuum. Ref.~\cite{Essler:2023nhu} showed that the 
finite-size scaling of off-diagonal matrix elements depends crucially on whether $|E_i\rangle$ and $|E_j\rangle$ (cf.~\eqref{eq:ETH}) are extracted from the same thermodynamic macrostate, i.e., statistical ensemble. 
Precisely, if $|E_i\rangle,|E_j\rangle$ are in the same macrostate, off-diagonal matrix elements decay exponentially 
as~\cite{Essler:2023nhu}  
\begin{equation}
\label{eq:fab1}
    \abs{\bra{E_i} \mathcal{O} \ket{E_j} } \propto \exp\!\left  [- c^{\mathcal{O}}\, L \ln(L)  - \widetilde{M}_{ij}^{\mathcal{O}}\, L\right ] ,
\end{equation}
where $c^{\mathcal{O}}>0$ is a constant that depends both on the operator $\mathcal{O}$ and on the 
macrostate. Moreover, the probability distribution function of 
$\widetilde{M}_{ij}^{\mathcal{O}}$ is well described by a Fr\'echet distribution.  
Curiously, the leading term in the large $L$ limit in the exponent of~\eqref{eq:fab1} exhibits a logarithmic correction.
\redd{Notice that the scaling~\eqref{eq:fab1} is similar to~\eqref{eq:ETH}, except for the logarithmic correction.
Moreover, 
we note that the Fr\'echet distribution differs from the distribution of the logarithm of the absolute value of a  Gaussian variable, implying that, unlike~\eqref{eq:ETH}, the matrix element $\bra{E_i}\mathcal{O}\ket{E_j}$ is non Gaussian.}
The matrix elements of local operators between eigenstates that are in 
different macrostates exhibit a faster decay in the thermodynamic limit as~\cite{Essler:2023nhu}  
\begin{equation}
 \label{eq:fab2}
    \abs{\bra{E_i} \mathcal{O} \ket{E_j}} \propto \exp\! \left (- d^{\mathcal{O}}\, L^2 \right ) ,
\end{equation}
where $d^{\mathcal{O}}>0$ depends on the macrostates from which $|E_i\rangle$ and $|E_j\rangle$ are 
extracted and on the operator. 
Eqs.~\eqref{eq:fab1} and~\eqref{eq:fab2} were supported by analytic calculations in the hard-core boson limit of the Lieb-Liniger model, and numerical results for the interacting case obtained by exploiting integrability~\cite{caux2007one,piroli2015exact}.

Although it is natural to expect that the scenario of Ref.~\cite{Essler:2023nhu} applies to generic integrable systems, 
several aspects deserve further investigation. For instance, Ref.~\cite{Essler:2023nhu} focused on the Lieb-Liniger model, which is a $1D$ quantum field theory model. The question whether the same scenario  applies to 
 integrable \emph{lattice} models, such as spin chains, has not been investigated yet. 
Moreover, Ref.~\cite{Essler:2023nhu} considered the \emph{repulsive} Lieb-Liniger model. Unlike most 
integrable systems, which possess multi-particle bound states (so-called ``strings'') of excitations, 
the repulsive Lieb-Liniger  model features only unbound excitations. Besides being interesting on their own, bound states are 
also experimentally observable (see Ref.~\cite{horvath2025observing} for a recent observation of ``strings'' in Bose gases). Also, typical equilibrium 
macrostates, such as thermal states, exhibit a finite density of bound states, implying that bound states are 
crucial to correctly describe equilibrium and out-of-equilibrium properties. Unfortunately, 
the presence of bound states renders the computation of matrix elements a daunting challenge. Indeed, 
despite the fact that exact formulas   
based on the Algebraic Bethe Ansatz  are available for the matrix elements of generic local operators~\cite{Kitanine:1998oii,Kitanine:2000srg}, they are plagued by fictitious 
singularities that have to be regularized to extract numerical results. 
The presence of bound states significantly increases the complexity of this regularization 
procedure.
Crucially, for operators that have support on a single site the singularities can be removed effectively~\cite{caux2005computation,caux2005computation1,caux2006dynamical,caux2009correlation}, and 
matrix elements  can be obtained 
with computational cost that grows only polynomially with system size, by exploiting the so-called string hypothesis~\cite{takahashi1999thermodynamics}. 
Some progress on the numerical computation of diagonal matrix elements of operators acting on more than one site 
was discussed in Ref.~\cite{Alba:2009th}. 
Precisely, it was shown that for eigenstates that do not contain strings it is possible 
to numerically construct the reduced density matrix of subsystems of length up to six contiguous sites. 
This approach was employed in Ref.~\cite{AlbaETHdiagonal} to investigate the \emph{ETH} scenario  for  
diagonal operators spin chains. 

Here we employ the strategy of Ref.~\cite{caux2005computation} and 
Ref.~\cite{Alba:2009th} to investigate the \emph{ETH} scenario for off-diagonal matrix elements of local operators having support on at most 
two sites in the spin-$1/2$ isotropic Heisenberg chain. Following Ref.~\cite{Essler:2023nhu}, we 
focus on eigenstates in the thermal macrostate (Gibbs ensemble). This is a natural choice because thermal states at a given energy density (temperature) are the most abundant ones, the number of non-thermal eigenstates being exponentially suppressed in the thermodynamic limit. We first focus on matrix elements between eigenstates in the same macrostate, considering one-spin operators. We show that off-diagonal matrix elements decay exponentially with $L$ similar to~\eqref{eq:fab1}.
This means that the presence of bound states does not alter the qualitative scenario of Ref.~\cite{Essler:2023nhu}. Next, we 
consider matrix 
elements between eigenstates in different macrostates. Precisely, we focus on the zero-temperature and 
the infinite-temperature macrostates.  Similar to~\eqref{eq:fab2}, off-diagonal matrix elements 
exhibit the faster decay as $e^{-L^2}$. 

We also numerically extract the probability distribution functions (\emph{PDF}) of $M_{ij}$ 
\begin{equation}
\label{eq:mij-intro}
M_{ij}=\ln|\langle E_i|\mathcal{O}|E_j\rangle|^2. 
\end{equation}
Quite generically, the \emph{PDF}s \redd{of the logarithm $M_{ij}$ of the absolute value of the matrix elements} are well described by Gumbel distributions, in contrast with Ref.~\cite{Essler:2023nhu}.
\redd{Similar to Ref.~\cite{Essler:2023nhu}, however, we observe that the fact that the distribution of $\widetilde M_{ij}$ is a Gumbel with $L$-independent parameters is incompatible with  the statistics of off-diagonal matrix elements $\langle E_i|\mathcal{O}|E_j\rangle$ being Gaussian (see \cref{sec:non-G} for numerical evidence).}
Next, we consider  two-spin operators. We focus on thermal macrostates constructed from eigenstates that do not contain strings. In contrast with one-spin operators, for which we can reach $L\sim 500$, for two-spin operators we only provide numerical data for $L\sim 60$. Despite that, our numerical data confirm the scenario observed for one-spin operators. 

The manuscript is organized as follows. In Section~\ref{sec:model} we introduce the Heisenberg chain, 
giving some details on the Bethe Ansatz treatment. In Section~\ref{sec:obs} we introduce the main observables 
that we consider, the thermodynamic macrostates, and their construction. In Section~\ref{sec:results} we 
discuss numerical results. We first consider the off-diagonal matrix elements between eigenstates in the same 
macrostate in Section~\ref{sec:same}. Specifically, we focus on one-spin operators in 
Section~\ref{sec:one-spin}, and on two-spin operators in Section~\ref{sec:two-spin}. In Section~\ref{sec:non-G} 
we show that the non Gaussianity of the \emph{PDF} describing off-diagonal matrix elements can be detected from  
the finite-size behavior of scale-invariant ratios constructed from~\eqref{eq:mij-intro}. This confirms 
early observations~\cite{leblond2019entanglement}. In Section~\ref{sec:2-macro} we discuss off-diagonal matrix 
elements between eigenstates in different macrostates. Finally, we conclude and discuss future directions 
in Section~\ref{sec:conclusions}. In Appendix~\ref{app:matrixElements} we provide the \emph{ABA} formulas for the 
matrix elements of local operators. Precisely, in Appendix~\ref{app:scalar} we detail the so-called Slavnov formula for 
the calculation of scalar products between arbitrary Bethe states. In Appendix~\ref{app:matrix} we discuss the  
formulas for the matrix elements of generic operators (in Appendix~\ref{app:gen-spin}) and for one-spin operators 
(in Appendix~\ref{app:one-spin}).

\section{Heisenberg spin chain}\label{sec:model}

Here we consider the spin-$1/2$ isotropic Heisenberg chain ($XXX$ chain), 
described by the Hamiltonian 
\begin{equation}\label{eq:XXXHam}
    H = \frac{J}{4} \sum_{i=1}^L \vec{\sigma}_{i} \cdot \vec{\sigma}_{i+1}\,,
\end{equation}
where $\vec\sigma_i=(\sigma_i^x,\sigma_i^y,\sigma_i^z)$ is the vector of Pauli matrices at 
site $i$, and $L$ is the length of the system. In the following we set $J=1$ in~\eqref{eq:XXXHam}. 
The Hamiltonian~\eqref{eq:XXXHam} possesses a $SU(2)$ symmetry generated by the total spin $\vec{S}_T = 1/2\sum_{i=1}^{L} \vec{\sigma}_i$, corresponding to global rotations of all the spins. 
As a consequence, both the total spin $\vec{S}^2_T$ and the total magnetization along the $z$ direction 
are good quantum numbers that can be used to label the different eigenstates. 

The $XXX$ chain~\cref{eq:XXXHam} is a paradigmatic example of Bethe Ansatz solvable many-body system~\cite{bethe1931zur,takahashi1999thermodynamics,korepin1993quantum,Essler2005the}. 
In the Bethe ansatz treatment of the $XXX$ chain the ferromagnetic state with all the spin up is 
a ``vacuum'' state, whereas overturned spins are treated as quasiparticle excitations. Each quasiparticle 
is labeled by a complex number $\lambda_i$, which is called rapidity. Any eigenstate in the sector with 
fixed number $M$ of quasiparticles, i.e., fixed magnetization, is identified by a set 
of $M$ rapidities $\lambda_i$. The rapidities satisfy nontrivial quantization conditions, encoded in 
a set of nonlinear coupled algebraic equations as 
\begin{equation}\label{eq:BetheEq}
    d(\lambda_i) \prod_{j\neq i} \frac{\lambda_j-\lambda_i-i}{\lambda_j-\lambda_i+i} = 1, 
\end{equation}
where the function $d(\lambda)$ is defined as
\begin{equation}\label{eq:ddef}
    d(\lambda) = \left (\frac{\lambda-\frac{i}{2}}{\lambda+\frac{i}{2}} \right )^{\!L}.
\end{equation}
The Bethe equations~\eqref{eq:BetheEq} admit $2^L$ sets of  solutions containing both real and complex rapidities.  
Complex rapidities correspond to bound states of quasiparticles. 
Although it is in general challenging to obtain the complex solutions of~\eqref{eq:BetheEq},  
it was already recognized by Bethe~\cite{bethe1931zur} that 
in the thermodynamic limit $L\to\infty$  some crucial simplifications occur. Precisely, 
the vast majority of complex rapidities form ``simple'' structures in the complex plane known as Bethe strings. Solutions of~\eqref{eq:BetheEq} in the same string have the same real part (string center) and imaginary parts that differ by  $\eta = \ii$. A string of length $n$ corresponds to a bound states of $n$ elementary excitations. In generic integrable systems, strings of arbitrary length are allowed.  

Within the framework of the string hypothesis, let us denote the rapidities 
as $\lambda^{j}_{n,\gamma}$, with $\gamma$ labeling the different strings of size $n$, i.e., that 
have different string centers, and $j$ labeling the rapidities in the same string. For large $L$ one has 
\begin{equation}
\label{eq:st-hyp}
\lambda_{n,\gamma}^{j}=\lambda_{n,\gamma}-i(n-2j-1)+i\delta^{j}_{n,\gamma}, 
\end{equation}
where $\delta_{n,\gamma}^j$ are the so-called string deviations, which account for the fact that 
the string hypothesis holds only in the limit $L\to\infty$. In~\eqref{eq:st-hyp} $\lambda_{n,\gamma}$ is the 
string center, and it is real. Real solutions of the Bethe equations correspond to $n=1$. 
For the vast majority of the eigenstates 
of the $XXX$ chain string deviations are suppressed exponentially 
as $\delta_{j,\gamma}^j=\mathcal{O}(e^{-\alpha L})$ (see Ref.~\cite{hagemans2007deformed} for a 
detailed study of string deviations in the $XXX$ chain). In particular, string 
deviations are not expected to affect the thermodynamic behavior of the model. By employing~\eqref{eq:st-hyp} 
in~\cref{eq:BetheEq}, after taking the limit $\delta_{n,\gamma}^j\to0$ and considering the logarithm of 
both members of the Bethe equations, one can write a set of equations 
for the string centers $\lambda_{n,\gamma}$. One obtains the so-called Bethe-Gaudin-Takahashi equations~\cite{takahashi1999thermodynamics} as
\begin{multline}
\label{eq:bgt}
L\theta_n(\lambda_{n,\gamma})=2\pi I_{n,\gamma}+\\\sum_{(m,\beta)\ne(n,\gamma)}\Theta_{m,n}(\lambda_{n,\gamma}-\lambda_{m,\beta}), 
\end{multline}
where we defined the scattering phases between different string types $\Theta_{n,m}$ as 
\begin{equation}
\label{eq:T1}
\Theta_{n,m}=\theta_{|n-m|}+\!\!\!\!\!\!\sum_{r=1}^{(n+m-|n-m|-1)/2}\!\!\!\!\!\!\!\!\!\!\!\!\!2\theta_{|n-m|+2r}+\theta_{n+m},
\end{equation}
for $n\ne m$, and as 
\begin{equation}
\label{eq:T2}
\Theta_{n,m}=\sum_{r=1}^{n-1}2\theta_{2r}+\theta_{2n},
\end{equation}
if $n=m$. In \cref{eq:bgt,eq:T1,eq:T2} we 
defined $\theta_n(x)=2\arctan(x/n)$. Let us denote by $s_n$ the number of 
strings of length $n$ in the eigenstate. Thus, one has the total number of particles as 
$M=\sum_{j=1}^M j s_j$. 
In~\eqref{eq:bgt}, $I_{n,\gamma}$ are the Bethe-Gaudin-Takahashi (\emph{BGT}) quantum numbers, and 
are integer or half-integer numbers for $L-s_n$ odd or even, respectively. For the $XXX$ chain 
the \emph{BGT} quantum numbers satisfy the bound as~\cite{takahashi1999thermodynamics}  
 \begin{equation}
 \label{eq:bound}
 |I_{n,\gamma}|\le \frac{1}{2}\big(L-1-\sum_{m=1}^M t_{m,n}s_m\big),
 \end{equation}
 with $t_{m,n}=2\min(m,n)-\delta_{n,m}$. In contrast with the Bethe equations~\eqref{eq:BetheEq}, 
 the \emph{BGT} equations are much simpler to solve, since their solutions are real.
 
 Moreover, by varying the $I_{n,\gamma}$ allowed by~\eqref{eq:bound}, one can target one by one the different solutions, i.e., the different eigenstates 
 of the $XXX$ chain. In conclusion, a 
 generic eigenstate of the $XXX$ chain is identified  by a string configuration $|s_1,s_2,\dots,s_M\rangle$, 
 with $s_j$ the number of strings of length $j$, and by the associated set of solutions of~\eqref{eq:bgt} as 
 \begin{equation}
 \label{eq:state}
 |\pmb{\lambda}\rangle=|\{\lambda_{j,\gamma_j}\}_{\gamma_j=1}^{s_j}\rangle_{j=1,2,\dots,M},
 \end{equation}
where, again, $\lambda_{j,\gamma_j}$ are the string centers for the strings of length $j$. 
It is important to stress that since the Hamiltonian~\eqref{eq:XXXHam} commutes with the total spin, its 
eigenstates are organized into $SU(2)$ multiplets, labeled by the eigenvalue of the 
total spin $S^2_{T}$ and the total magnetization $S^z_{T}$. Starting from an eigenstate with  
a given $S_{T}$ and $S_T^z=S_T$, i.e., with maximum value of the magnetization (highest-weight state), all the descendent  
eigenstates in the multiplet are obtained by repeated applications of the spin lowering 
operator $S^-_T=\sum_{j=1}^L S^-_j$, with $S_j^-=(\sigma_j^x-i\sigma_j^y)/2$. 
One should stress that by solving Eq.~\eqref{eq:state} for a given $M$, one obtains only the eigenstates with  
$S_T^z=S_T$. However, it is well-known~\cite{takahashi1999thermodynamics} that each application of the 
lowering operator $S^-_T$ corresponds to adding one extra infinite rapidity to the solution of the Bethe equations. 
 
Finally, after solving~\eqref{eq:bgt} for the \emph{BGT} rapidities $\lambda_{n,\gamma}$, one can easily obtain  
the eigenstate expectation value of all the conserved quantities. For instance, the energy reads as 
\begin{equation}
\label{eq:ener}
E=-\sum_{j,\gamma_j}\frac{2j}{\lambda_{j,\gamma_j}^2+j^2}.
\end{equation}
However, here we are interested in matrix elements of local operators, which are not straightforwardly obtained from the solutions of~\eqref{eq:bgt}. Indeed, the eigenfunctions of the $XXX$ chain, which would allow to compute the matrix elements, contain ${\mathcal O}(2^L)$ components, and are impractical. Moreover, the string hypothesis 
leads to fictitious singularities in the wavefunction components that have to be carefully  removed to extract 
numerical data. Instead of working with the eigenstates wavefunctions, here we will exploit the 
fact that exact formulas for the matrix elements have been obtained by using the 
\emph{ABA} approach~\cite{korepin1993quantum} (see Section~\ref{sec:obs}).  

\subsection{Macrostates and observables}
\label{sec:obs}

Here we are interested in the statistics of off-diagonal matrix elements of local operators 
in the thermodynamic limit $L,M\to\infty$ with $M/L$ fixed. To characterize the thermodynamic behavior of 
local observables, it is convenient to work with macrostates rather than with 
individual eigenstates. Macrostates are ensembles of eigenstates that have the same local properties in the 
thermodynamic limit. Macrostates are identified by the expectation value of \emph{extensive} conserved 
quantities. For instance, by fixing only the value of the energy density $E/L$, one obtains the 
thermal macrostate at a given inverse temperature $\beta$. 
The thermal macrostate describes the ``typical'' eigenstates at a given energy density.
Indeed, although for each value of $E/L$ in the limit $L\to\infty$ there is an exponentially diverging number of eigenstates 
that are atypical, i.e., giving nonthermal expectation values for local observables, 
their fraction is exponentially vanishing compared to the thermal-like eigenstates, which 
are exponentially more abundant.
Notice that the presence of an exponentially diverging number of 
atypical eigenstates gives rise to nonthermal steady states after quantum quenches in 
integrable systems, whose description is provided by the Generalized Gibbs Ensemble~\cite{calabrese2016introduction} (\emph{GGE}). 
On the other hand, for nonintegrable systems atypical eigenstates are expected to be suppressed~\cite{biroli2010effect}. 

Following Ref.~\cite{AlbaETHdiagonal} and Ref.~\cite{Essler:2023nhu}, here 
we consider matrix elements between eigenstates extracted from thermal macrostates at temperature 
$\beta$. Precisely, we consider  the thermal density matrix $\rho_{th}$  
defined as 
\begin{equation}
\label{eq:rho-the}
\rho_{th}(\beta)=Z^{-1}\sum_{|\pmb{\lambda}\rangle} e^{-\beta E(\pmb{\lambda})}|\{\pmb{\lambda}\}\rangle
\langle\{\pmb{\lambda}\}|, 
\end{equation}
where $Z$ is the partition function. As stressed above, in the thermodynamic limit 
all the eigenstates in~\eqref{eq:rho-the} give the same expectation values of local observables. 
This is the content of the so-called \emph{weak} Eigenstate Thermalization Hypothesis~\cite{biroli2010effect}, and it is 
also at the heart of the Thermodynamic Bethe Ansatz ($TBA$) approach for integrable systems~\cite{takahashi1999thermodynamics}.  
Now, since the number of eigenstates in~\eqref{eq:rho-the} grows exponentially with $L$, 
and the sum in~\eqref{eq:rho-the} cannot be performed explicitly,  we adopt a sampling strategy, as in 
Ref.~\cite{Essler:2023nhu} and~\cite{AlbaETHdiagonal}. Precisely, we employ a Metropolis scheme as in Refs.~\cite{alba2015simulating,alba2016the,AlbaETHdiagonal} to sample the eigenstates of the $XXX$ chain 
with the Gibbs weight $e^{-\beta E}/Z$. Moreover, we restrict ourselves to 
fixed number of particles, i.e., fixed ratio $M/L$. We only consider the eigenstates that 
do not contain infinite rapidities, i.e., eigenstates with  largest magnetization (compatibly with $M$). 
It is natural to expect that these restrictions do not alter the qualitative scenario, similar to 
what observed in Ref.~\cite{AlbaETHdiagonal}. 

Here we consider local observables  that are built from  
the operators $E_i^{(mn)}$ localized on site $i$, and defined as 
\begin{equation}\label{eq:elementary}\begin{aligned}
    &E_i^{(11)} = \frac{1}{2}(\mathds{1}_i+\sigma^z_i),
    \quad
    &&E_i^{(12)} = 
  \frac{1}{2}(\sigma^x_i+i\sigma^y_i),\\
    &E_i^{(21)} = 
    \frac{1}{2}(\sigma_i^x-i\sigma_i^y),
    &&E_i^{(22)} 
    = \frac{1}{2}\left ( \mathds{1}_i-\sigma^z_i \right ).
\end{aligned} 
\end{equation}
In the following section, in particular, we will study the statistics of 
the off-diagonal matrix elements of $E_i^{(11)}$ and $E_i^{(11)} E_{i+1}^{(22)}$, taken 
as prototypical 
examples of one and two-spin operators, respectively. 
The operators~\eqref{eq:elementary} are particularly convenient to compute in the $XXX$ model~\eqref{eq:XXXHam}, by virtue of the exact results of Refs.~\cite{Kitanine:1998oii,Kitanine:2000srg}.
Indeed, a major breakthrough~\cite{Kitanine:1998oii,Kitanine:2000srg} is that 
 it is possible to rewrite the  operators~\eqref{eq:elementary} in terms of the matrix elements of the 
so-called transfer matrix of the $XXX$ chain~\cite{korepin1993quantum}, which is at the heart of the 
\emph{ABA} solution of the model.  
Then, by using the known action of the transfer matrix on the eigenstates of~\eqref{eq:XXXHam}, 
it is possible to obtain the matrix elements. We review the main results of 
Refs.~\cite{Kitanine:1998oii,Kitanine:2000srg} in Appendix~\ref{app:matrixElements}, 
reporting the explicit expression for the matrix elements of the operators in~\eqref{eq:elementary}. 
Unlike exact diagonalization approaches, the computational cost to evaluate  
matrix elements via the \emph{ABA} results 
reported in~\cref{app:matrixElements} scales exponentially only in the size $\ell$ of the support of the 
operator (here we restrict ourselves to 
$\ell = 1, 2$), but polynomially as $O(M^\ell)$ in the number of particles $M$. This allows us to reach chain 
sizes $L\sim 500$ which are prohibitively large for exact diagonalization. 
Still, we should mention that the numerical implementation of the \emph{ABA} formulas 
for the matrix elements is nontrivial, because they are derived for an inhomogeneous 
version of the $XXX$ chain. The homogeneous limit that is  required  to 
recover~\eqref{eq:XXXHam} introduces fictitious singularities that have to be removed. 
Furthermore, the presence of string solutions introduces additional singularities, which 
have  to be dealt with by carefully taking the limit of vanishing string deviations. 
For the operators with support on more than one site, treating the homogeneous 
limit and the limit of vanishing string deviations at the same time is very challenging.
For this reason, as in Ref.~\cite{Alba:2009th}, we only consider off-diagonal matrix elements 
of {$E_i^{(11)} E_{i+1}^{(22)}$} between eigenstates without bound states. 
For one-spin operators, since the homogeneous limit is trivial, 
the limit of vanishing string deviations can be performed analytically.
Thus, for one-spin operators we 
are able to investigate the effect of bound states on the finite-size scaling of 
matrix elements.

\section{Numerical results}\label{sec:results}

\begin{figure}    
    \centering
    \includegraphics[width=\linewidth]{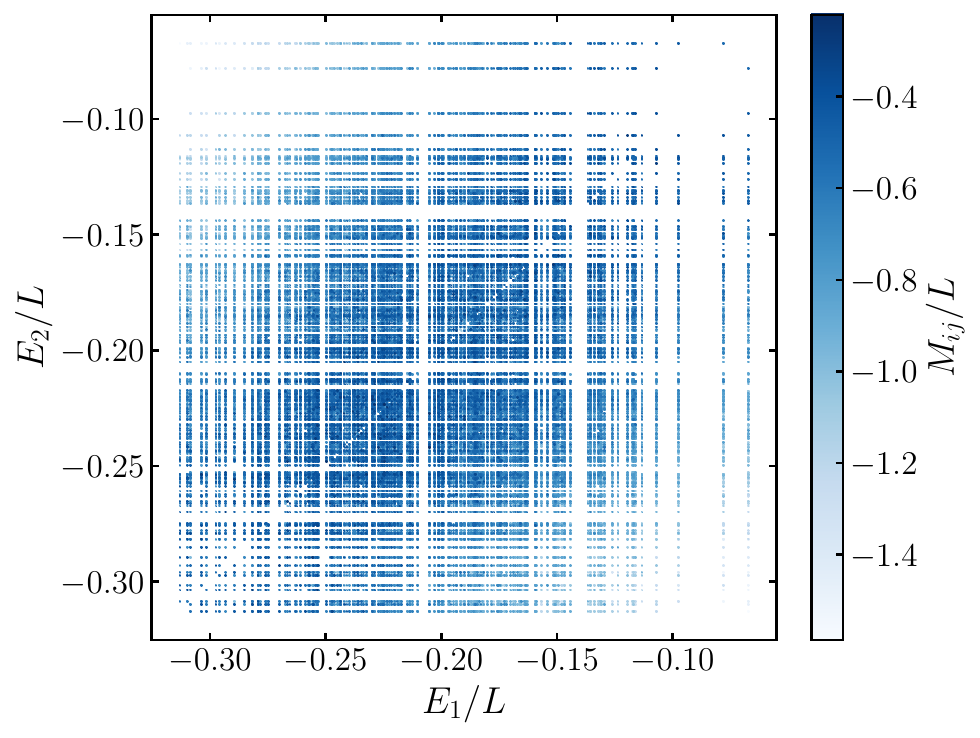}
    \caption{Logarithm of the matrix elements for the one-spin operator $E^{(11)}_j$
    with $L = 56,\, M = 14$ as a function of the energy densities $E_i/L$ of the two 
    eigenstates (diagonal elements not reported).}
    \label{fig:matrices}
\end{figure}

\begin{figure}
    \centering
    \includegraphics[width=\linewidth]{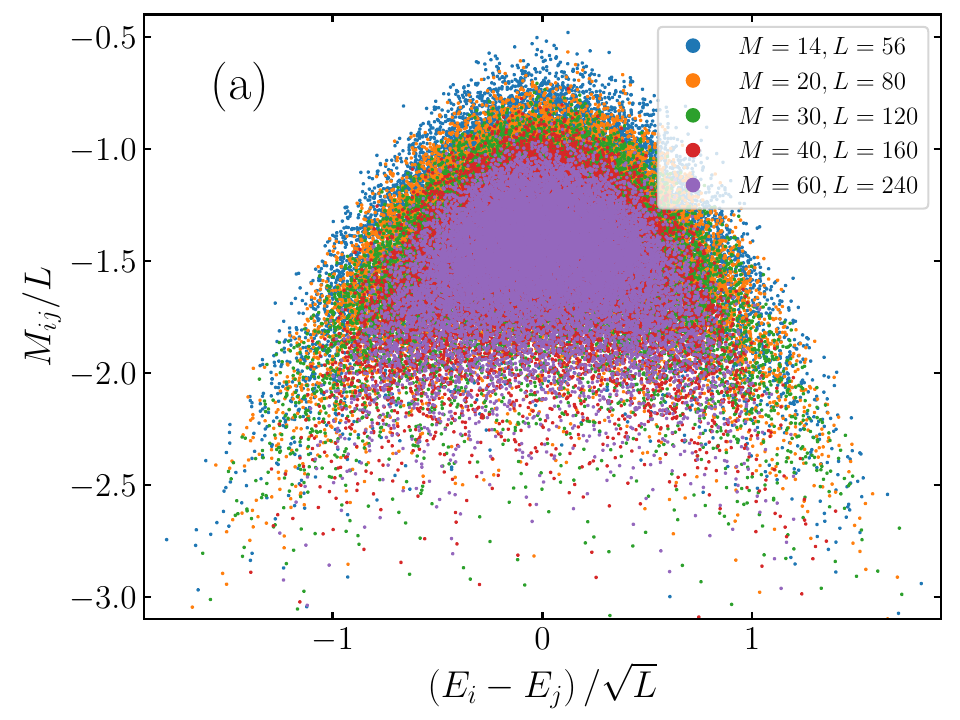}
    \includegraphics[width=\linewidth]{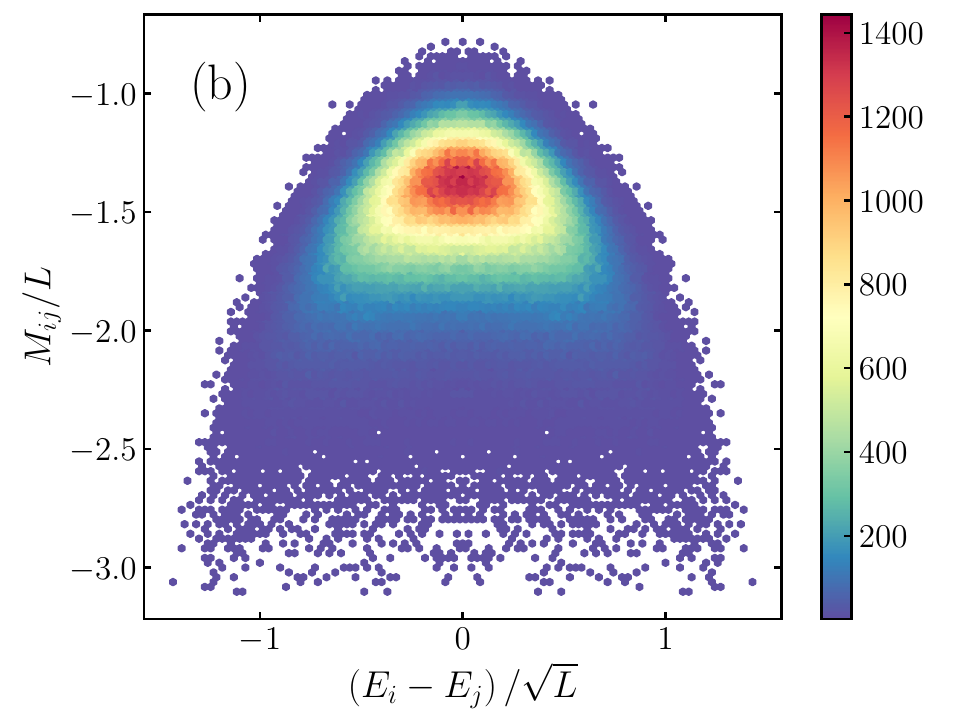}
    \caption{Scaling of the logarithm $M_{ij}$ (cf.~\cref{eq:Mdef}) of the matrix elements of the 
    one-spin operator $E_i^{(11)}$ as a function of $E_i-E_j$, with $E_i,E_j$ the energy of the 
    two eigenstates.
    (a) Scatter plot of the logarithm $M_{ij}$ rescaled by $L$ as a function of $(E_i-E_j)\sqrt{L}$ 
    for $L = 56, 80, 120, 160,$ and $240$ and fixed $M = L/4$.
    (b) Two-dimensional  histogram of the rescaled logarithm $M_{ij}/L$ as a function of $(E_i - E_j)/\sqrt{L}$.
    }
    \label{fig:scatter}
\end{figure}
Before discussing our main results, let us discuss some qualitative features of 
off-diagonal matrix elements. 
First, given the operator $\mathcal{O}$ of interest, let us define the main 
quantity of interest $M_{ij}$ as 
\begin{equation}\label{eq:Mdef}
    M_{ij} = \ln \abs{\bra{E_i}\mathcal{O}\ket{E_j}}^2 = \ln \abs{\mathcal{O}_{ij}}^2.
\end{equation}
In \cref{fig:matrices} we overview the behavior of $M_{ij}$ for the 
one-spin operator $E^{(11)}_i$ (cf.~\eqref{eq:elementary}) in the $XXX$ chain with 
$L = 56$, $M = L/4$. The data shown in the figure are for the infinite-temperature 
macrostate. Precisely, by employing the method of Ref.~\cite{alba2015simulating} we 
generate an ensemble of $\sim 10^3$ eigenstates of the $XXX$ chain distributed with 
the infinite-temperature Gibbs probability. As discussed earlier, we fix the magnetization 
$M/L=1/4$ and consider only the eigenstates that correspond to solutions of the Bethe 
equations without infinite rapidities. On the two axes we show the energy density $E_1/L$ and 
$E_2/L$ of the eigenstates. Here we have $E_1/L,E_2/L\in [-0.3,-0.1]$. 
Notice that in the thermodynamic limit $L,M\to\infty$, the fluctuations of the energy density 
are suppressed as $1/\sqrt{L}$, i.e., the energy density peaks at the thermal value at 
the density $M/L=1/4$. Again, this reflects that in the thermodynamic 
limit all the eigenstates in the thermal ensemble give the same expectation value for 
local observables. 
The dots in the plot are the values of $M_{ij}/L$, the color intensity being their 
magnitude. We omit the diagonal matrix elements. The smallest off-diagonal 
matrix elements are $\approx 10^{-25}\sim e^{-L}$. 

It is interesting to investigate the dependence of $M_{ij}$ on the energy difference 
$E_i-E_j$. This is illustrated in~\cref{fig:scatter}~(a) plotting $M_{ij}/L$ versus 
$(E_i-E_j)/\sqrt{L}$, where the $1/\sqrt{L}$ accounts for the fact that in the thermodynamic 
limit the energy peaks around its thermal value with $\sqrt{L}$ fluctuations. The different 
colors now correspond to different chain sizes $L\le 240$, with fixed particle density $M/L=1/4$. 
The data for different $L$ overlap in the large $L$ limit, suggesting that $M_{ij}\sim L$. The largest off-diagonal matrix 
elements $M_{ij}$ occur for $E_j=E_i$, and $M_{ij}$  increases if the eigenstates are farther apart in 
energy. In \cref{fig:scatter}~(b) we report with the color scale the number of 
matrix elements. Clearly, the most abundant values of $M_{ij}$ are at 
$E_i-E_j\approx 0$ and  peak around $M_{ij}/L\approx -1.5$.

\subsection{Statistics of matrix elements between eigenstates in the same macrostate}
\label{sec:same}

Here we quantitatively address the finite-size scaling of off-diagonal matrix elements and their 
statistics. Precisely, we consider both one-spin and two-spin operators in Section~\ref{sec:one-spin} 
and Section~\ref{sec:two-spin}, respectively. Finally, in Section~\ref{sec:non-G} we show that 
the non Gaussianity of the  probability distribution of off-diagonal matrix elements  
is reflected in the divergence with $L$ of an ad hoc defined ratio of matrix elements. 

\subsubsection{One-spin operator}
\label{sec:one-spin}

\begin{figure*}
    \centering
    \includegraphics[width=.32\linewidth]{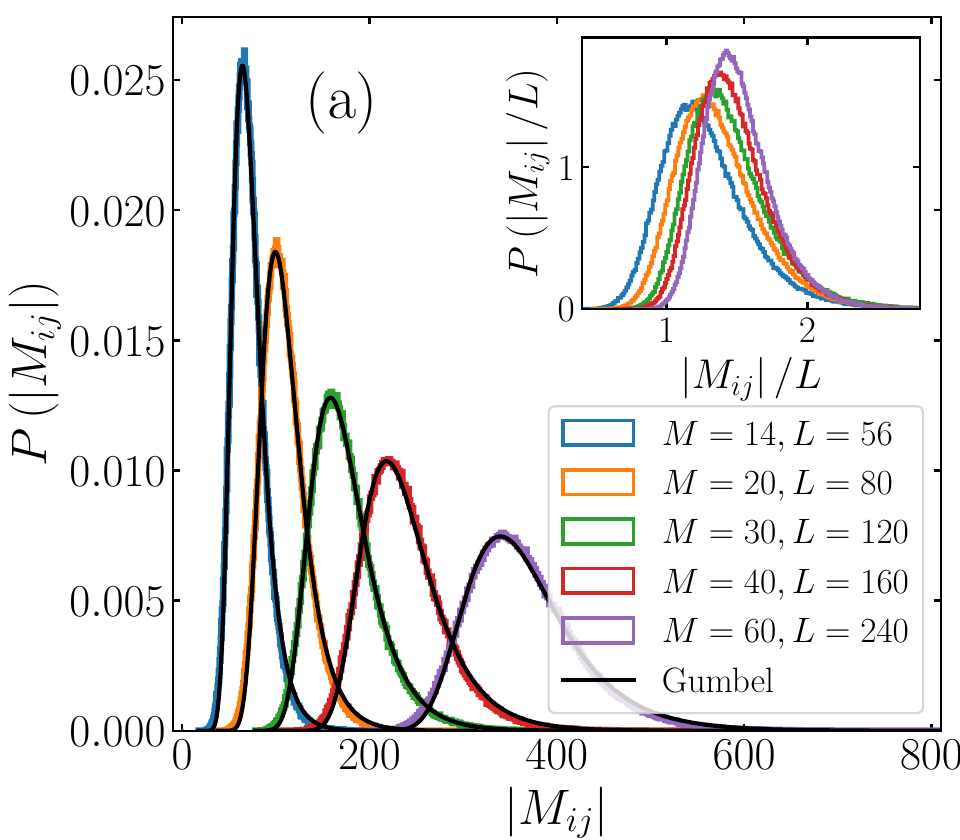}
    \includegraphics[width=.32\linewidth]{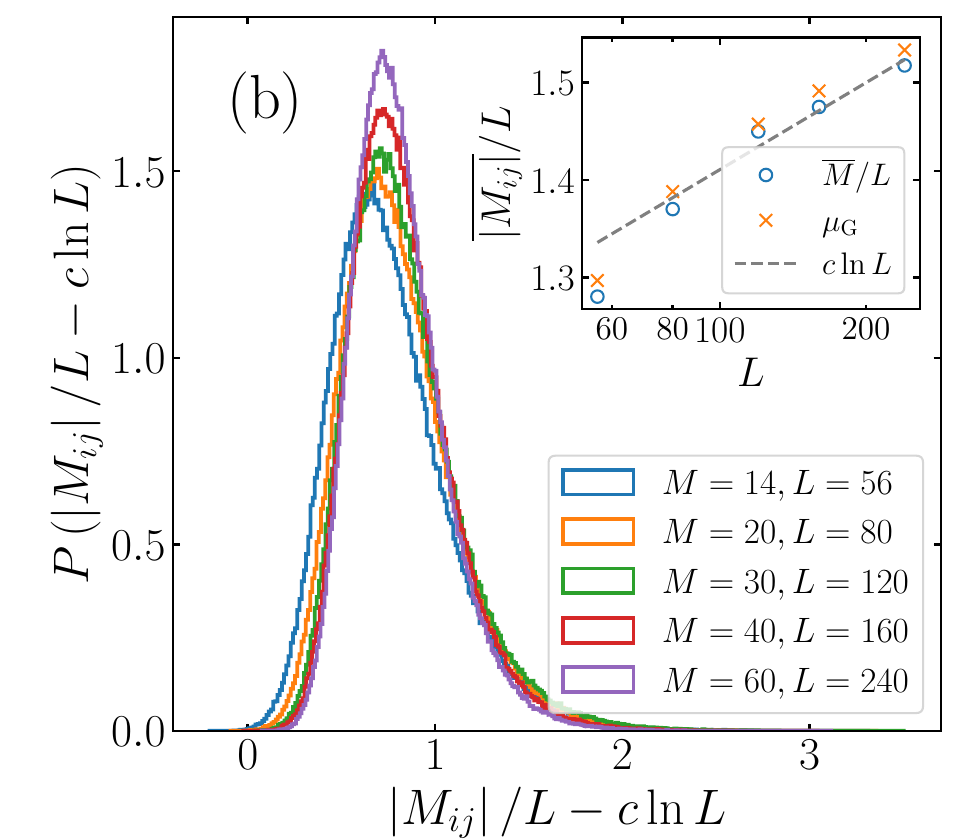}
    \includegraphics[width=.32\linewidth]{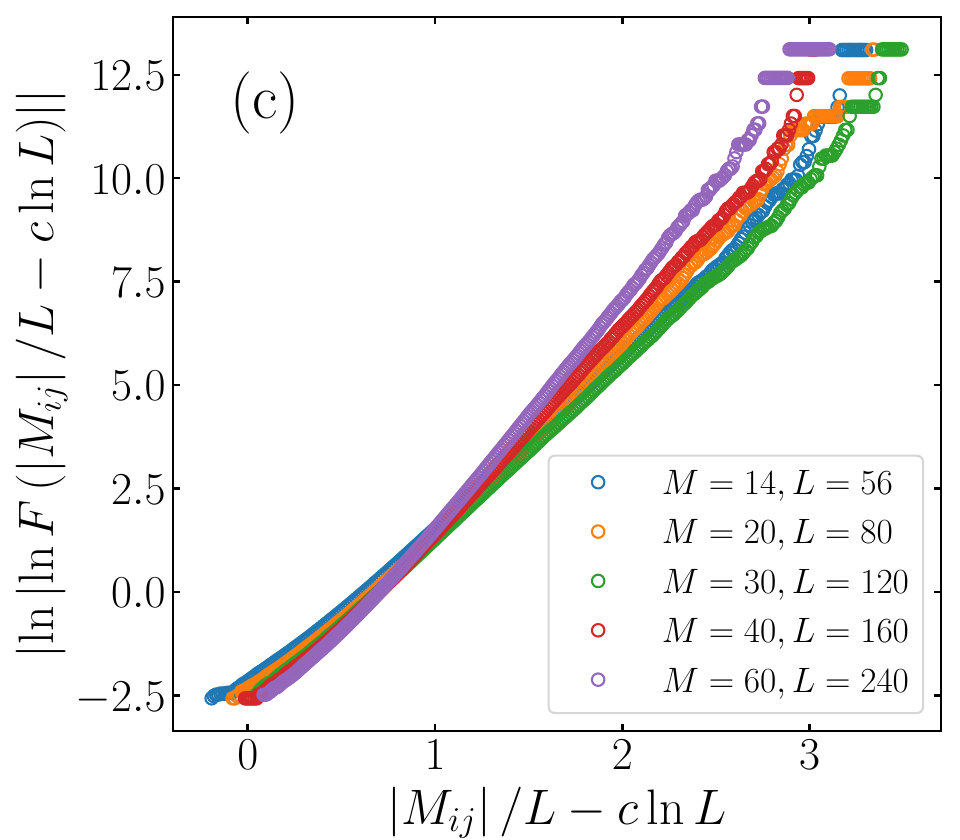}
    \caption{Distribution of the logarithm $\abs{M_{ij}}$ (cf.~\eqref{eq:Mdef}) of the matrix elements of the 
    one-spin operator $E_i^{(11)}$ (cf.~\eqref{eq:elementary}) for eigenstates of the $XXX$ chain with 
    particle number $M = 14, 20, 30, 40,$ and $60$ and chain size 
    $L=4M$. Eigenstates are extracted from the infinite-temperature thermal macrostate. 
    (a)~Histogram of $\abs{M_{ij}}$. The continuous lines are fits to the Gumbel distribution~\eqref{eq:Gumbel}, 
    with $\beta,\mu$ fitting parameters. 
    In the inset we report the histogram of $\abs{M_{ij}}/L$. 
    (b)~Histograms of the shifted $M_{ij}/L-c\ln(L)$, with $c$ a parameter  
    obtained by fitting the average $\overline{\abs{M_{ij}}}/L$ to the behavior $c\ln(L)+c_0$, with $c_0$ another 
    fitting parameter. The fit is reported as gray-dashed line in the inset. In the inset 
    $\mu_G$ is the mean of the Gumbel 
    distribution, i.e., $\mu_G=\mu+\beta \gamma$, with $\beta,\mu$ as 
    in~\eqref{eq:Gumbel} and $\gamma$ the Euler-Mascheroni constant. To compute $\mu_G$ we used the fitted 
    values of $\beta,\mu$ in panel (a). 
    (c)~Double logarithm of the cumulative distribution function (\emph{CDF}) $F(\abs{M_{ij}}/L - c\ln L)$ as extracted from 
    the numerical data. 
    }
    \label{fig:histogram_1spin_rat4_same_macro}
\end{figure*}

\begin{figure*}
    \centering
    \includegraphics[width=.32\linewidth]{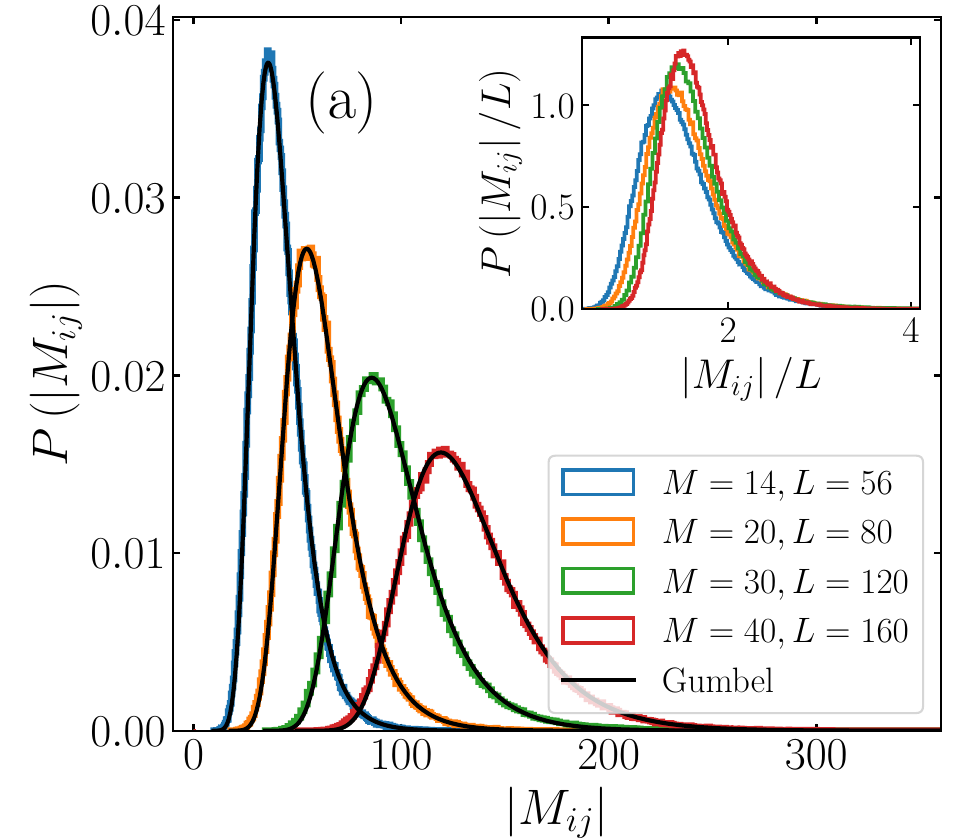}
    \includegraphics[width=.32\linewidth]{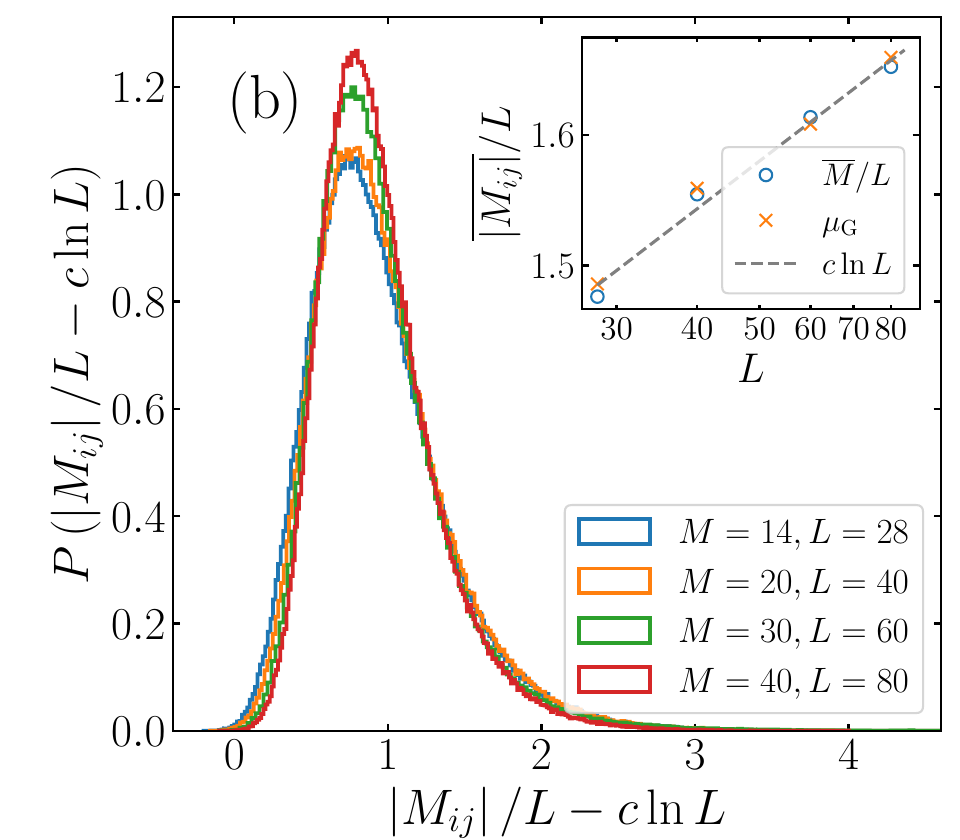}
    \includegraphics[width=.32\linewidth]{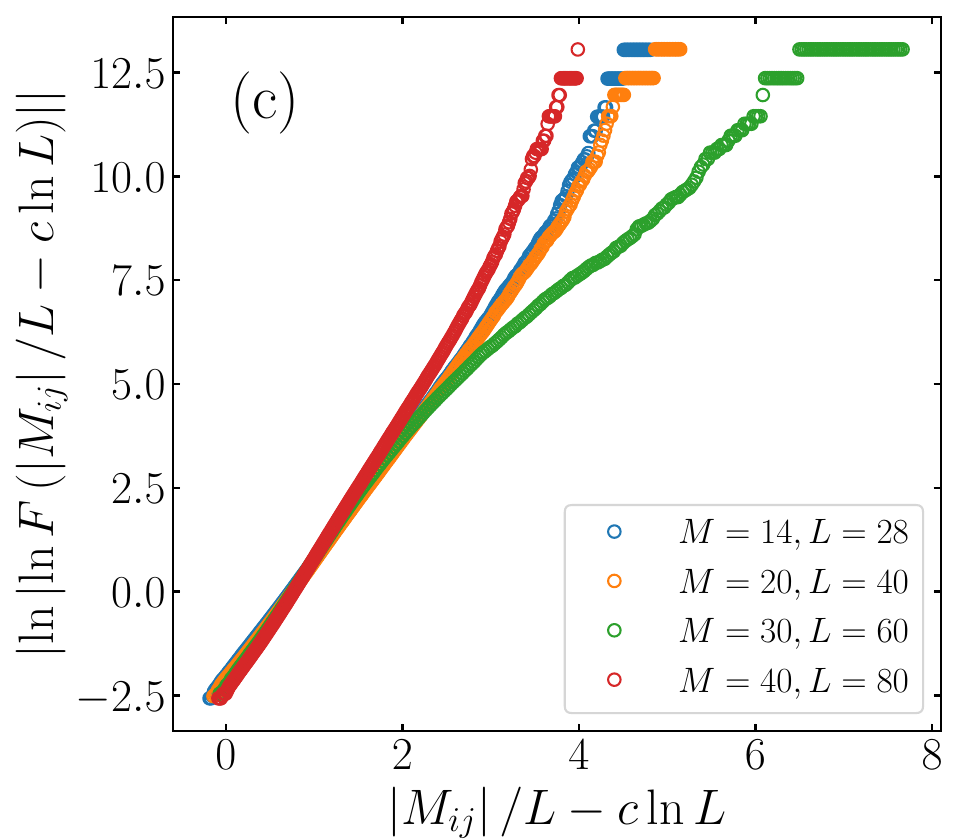}
    \caption{Same analysis as in Fig.~\ref{fig:histogram_1spin_rat4_same_macro} for eigenstates of the $XXX$ chain in 
    the infinite-temperature macrostate with particle density $M/L=1/2$. 
    }
    \label{fig:histogram_1spin_rat2_same_macro}
\end{figure*}

\begin{figure*}
    \centering
    \includegraphics[width=.32\linewidth]{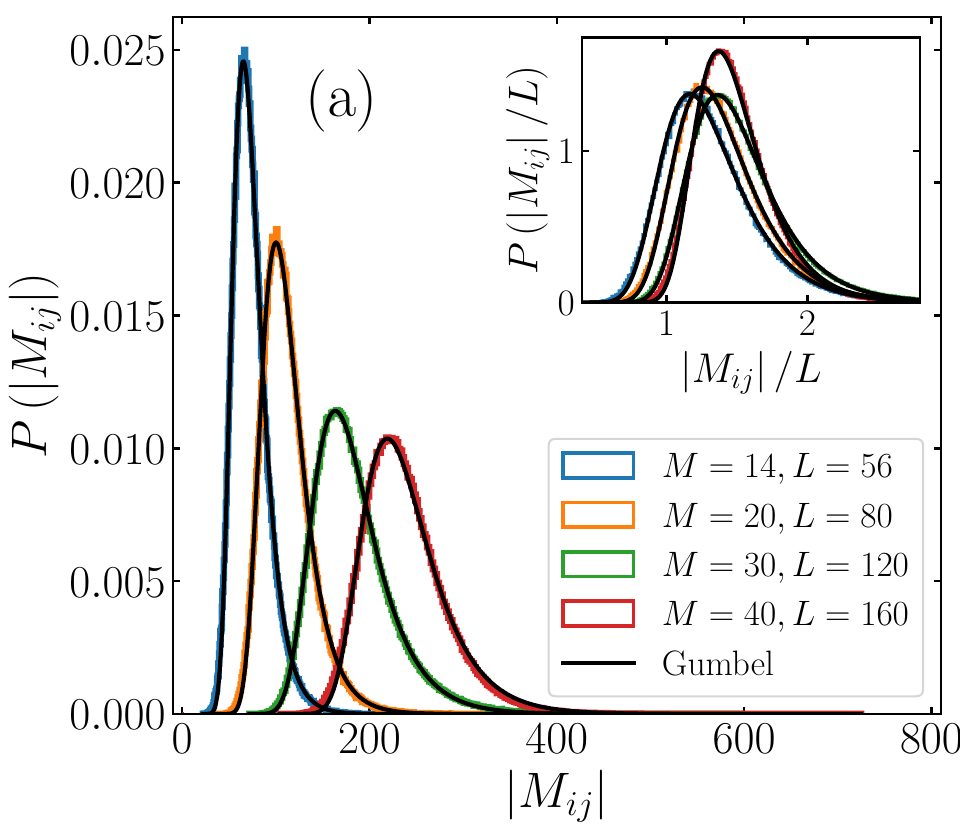}
    \includegraphics[width=.32\linewidth]{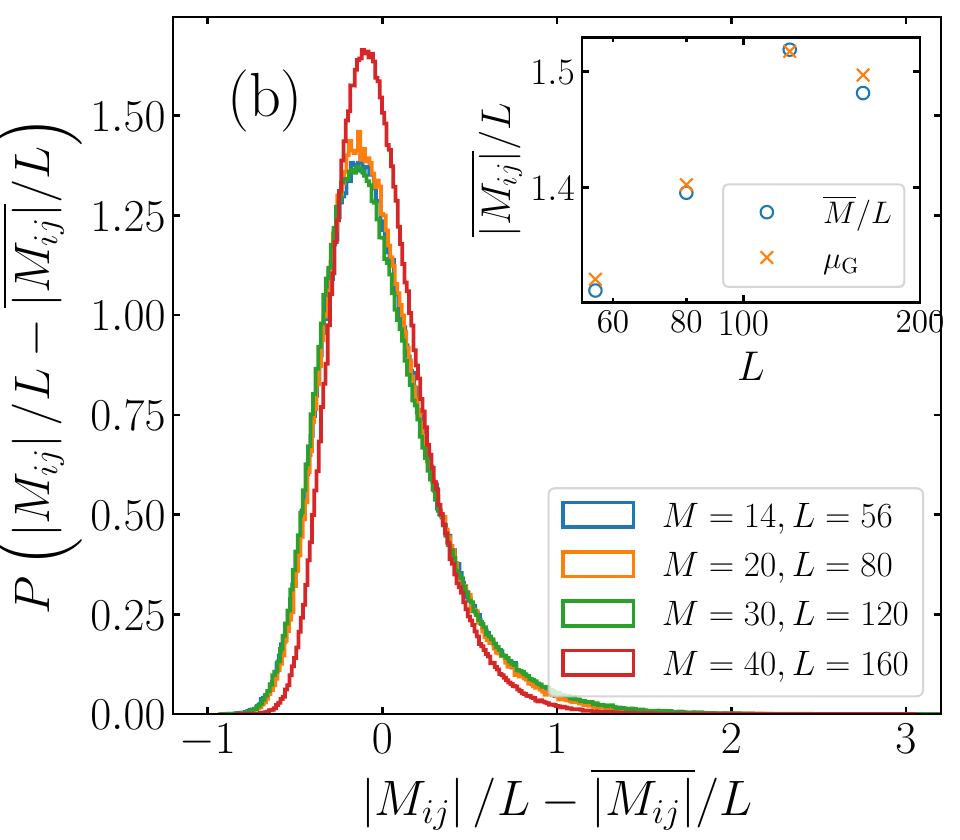}
    \includegraphics[width=.32\linewidth]{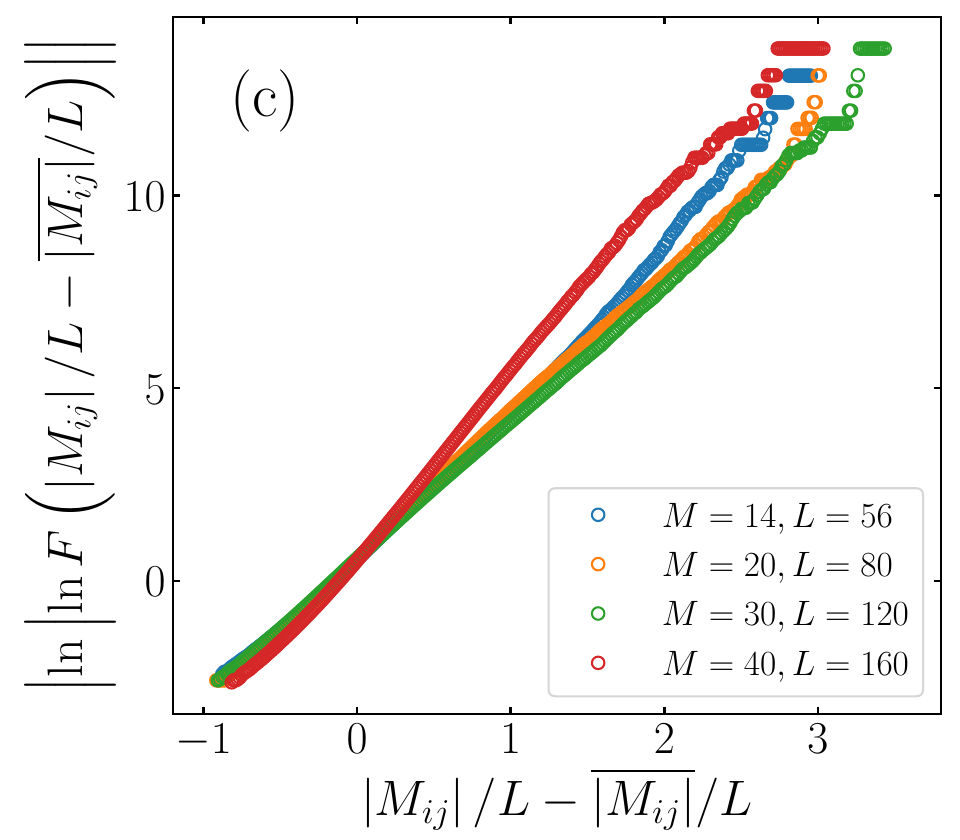}
    \caption{\fed{
    Distribution of $M_{ij}$~\eqref{eq:Mdef} of the one-spin operator $E_i^{(11)}$ in \cref{eq:elementary} for eigenstates of the $XXX$ chain in the finite temperature macrostate at $\beta = 0.5$ with particle density $M/L=1/4$ and particle numbers $M = 14, 20, 30$ and $40$.
    (a)~Histogram of $\abs{M_{ij}}$.
    The continuous lines are fits to the Gumbel distribution~\eqref{eq:Gumbel}.
    In the inset we report the histograms of $\abs{M_{ij}}/L$.
    (b)~Histograms of the shifted $\abs{M_{ij}}/L - \overline{\abs{M_{ij}}}/L$, where $\overline{\abs{M_{ij}}}$ is the average of $\abs{M_{ij}}$.
    In the inset we report the averages $\overline{\abs{M_{ij}}}$ and we compare them with the mean $\mu_G$ of the Gumbel distribution~\eqref{eq:Gumbel} obtained from the fits in panel (a).
    (c)~Double logarithm of the \emph{CDF} of $\abs{M_{ij}}/L - \overline{\abs{M_{ij}}}/L$.
    }
    }
    \label{fig:histogram_1spin_rat4_same_macro_beta0.5}
\end{figure*}

In~\cref{fig:histogram_1spin_rat4_same_macro} we report $M_{ij}$ (cf.~\eqref{eq:Mdef}) for 
the operator $E^{(11)}_i$ (cf.~\eqref{eq:elementary}). In~\cref{fig:histogram_1spin_rat4_same_macro}~(a) we show the 
histograms $P(|M_{ij}|)$ for several chain sizes up to $L=240$, at fixed 
particle density $M/L=1/4$. The data are the same as in \cref{fig:scatter}~(b) for the 
infinite-temperature macrostate. We observe that as $L$ increases, the 
center of the peak moves towards larger values of $|M_{ij}|$, and 
the distribution broadens. 
\redd{Moreover, the histograms are not symmetric around the maximum.}
Notice that at $L=240$ the typical off-diagonal matrix elements is  $M_{ij}\sim 10^{-200}$. 
Indeed, to correctly capture such tiny matrix elements we employed arbitrary-precision arithmetic 
in evaluating the \emph{ABA} formulas (see Appendix~\ref{app:matrixElements}). 
Let us now discuss the statistics of the off-diagonal matrix elements, i.e., the functional form 
of the distribution $P(M_{ij})$. For the Lieb-Liniger gas, Ref.~\cite{Essler:2023nhu} 
showed that $P(|{M}_{ij}|/L)$ is well described by a Fr\'echet 
distribution $P_{\alpha,\beta,\nu}$, 
whose probability distribution function takes the form
\begin{equation}
\label{eq:frechet}
P_{\alpha,\beta,\nu}(x)=
(x-\nu)^{-\alpha-1}\exp\Big[-\Big(\frac{x-\nu}{\beta}\Big)^{-\alpha}\Big],
\end{equation}
for $x>\nu$, and it is zero otherwise. By fitting the data in Fig.~\ref{fig:histogram_1spin_rat4_same_macro}  to~\eqref{eq:frechet}, we obtain $\beta\approx-\nu$ and quite large 
values of $\beta\sim 10^3$. In the limit of large $\beta=-\nu$ 
the Fr\'echet distribution reduces to the Gumbel distribution $P_{\mu,\beta}$ 
given by 
\begin{equation}\label{eq:Gumbel}
    P_{\mu,\beta}(x) = \frac{1}{\beta} \, e^{-\frac{x-\mu}{\beta}-e^{-\frac{x-\mu}{\beta}}}.
\end{equation}
The continuous lines in Fig.~\ref{fig:histogram_1spin_rat4_same_macro} are fits to~\eqref{eq:Gumbel} 
with $\beta,\mu$ the fitting parameters. For all the values of $L$ the data are quite well described 
by~\eqref{eq:Gumbel}. 
To proceed, we observe that according to 
\emph{ETH} (cf.~\eqref{eq:ETH}), one should expect that $M_{ij}\sim -L$, which has been verified in Ref.~\cite{Essler:2023nhu} 
for the Lieb-Liniger model. Thus, in the inset of \cref{fig:histogram_1spin_rat4_same_macro} we report the histograms $P(M_{ij}/L)$. The collapse of the data for different $L$ is not perfect, as a residual ``drift'' 
towards the right is visible. This is expected, and it was observed also for the Lieb-Liniger gas~\cite{Essler:2023nhu}. This effect  
can be fully accounted for by a residual $L$ dependence of the parameter $\mu$ of the distribution, which is 
related to the average of the Gumbel distribution $\mu_G=\mu+\beta\gamma$, with $\gamma$  the Euler-Mascheroni constant. 
Thus, let us consider the average $\overline{M_{ij}}$ as a function of $L$. 
Following Ref.~\cite{Essler:2023nhu} we fit $\overline{M_{ij}}/L$ to the behavior 
\begin{equation}
\label{eq:fit}
\overline{M_{ij}}=-cL\ln(L)-c_0 L, 
\end{equation}
with $c,c_0$ fitting parameters. By fitting the data with $L\ge 80$ we obtain 
$c \approx 0.13$ and $c_0 \approx  0.82$. Now,  
the histograms of $|M_{ij}|-c\ln(L)$ for different values of $L$ 
should collapse on the same curve, at least in the limit $L\to\infty$. 
In \cref{fig:histogram_1spin_rat4_same_macro}~(b) we show the distribution of $|M_{ij}|/L-c\ln(L)$, 
with $c$ the fitted value. The data collapse for different $L$ 
improves as compared with Fig.~\ref{fig:histogram_1spin_rat4_same_macro}~(a), 
although it is not perfect. This could be attributed to residual finite $L$ corrections. 
In the inset of \cref{fig:histogram_1spin_rat4_same_macro}~(b) we report  
 $|\overline{M_{ij}}|/L$ versus $L$ (notice the logarithmic scale on the $x$-axis). 
 The dashed line is the fit to~\eqref{eq:fit}. The crosses denote the average of the 
 Gumbel distribution $\mu_G$ computed from the fitted constants $\mu,\beta$ extracted in 
 panel (a), and are compatible with $\overline{M}_{ij}/L$. 
It is also interesting to consider the Cumulative Distribution Function (\emph{CDF}).  
The \emph{CDF} $F(x)$ of the Gumbel distribution is 
\begin{equation}\label{eq:GumbelCDF}
    F(x) = \int_{-\infty}^x P_{\mu,\beta}(t)dt= e^{- e^{-\frac{x-\mu}{\beta}}}.
\end{equation}
Eq.~\eqref{eq:GumbelCDF} implies that the double logarithm of $F(x)$ is a straight line. 
In \cref{fig:histogram_1spin_rat4_same_macro}~(c) we show  the double logarithm of the \emph{CDF} of 
$\abs{M_{ij}}/L - c\ln(L)$ as extracted from the numerical data. 
The data 
exhibits an approximate linear behavior as a function $|M_{ij}|/L-c\ln(L)$ in the central region, whereas  
deviations from linear scaling are visible at the edges, which correspond to the tails of the distribution,  
where finite $L$ effects are expected to be larger. 

\begin{figure*}
    \centering
    \includegraphics[width=.32\linewidth]{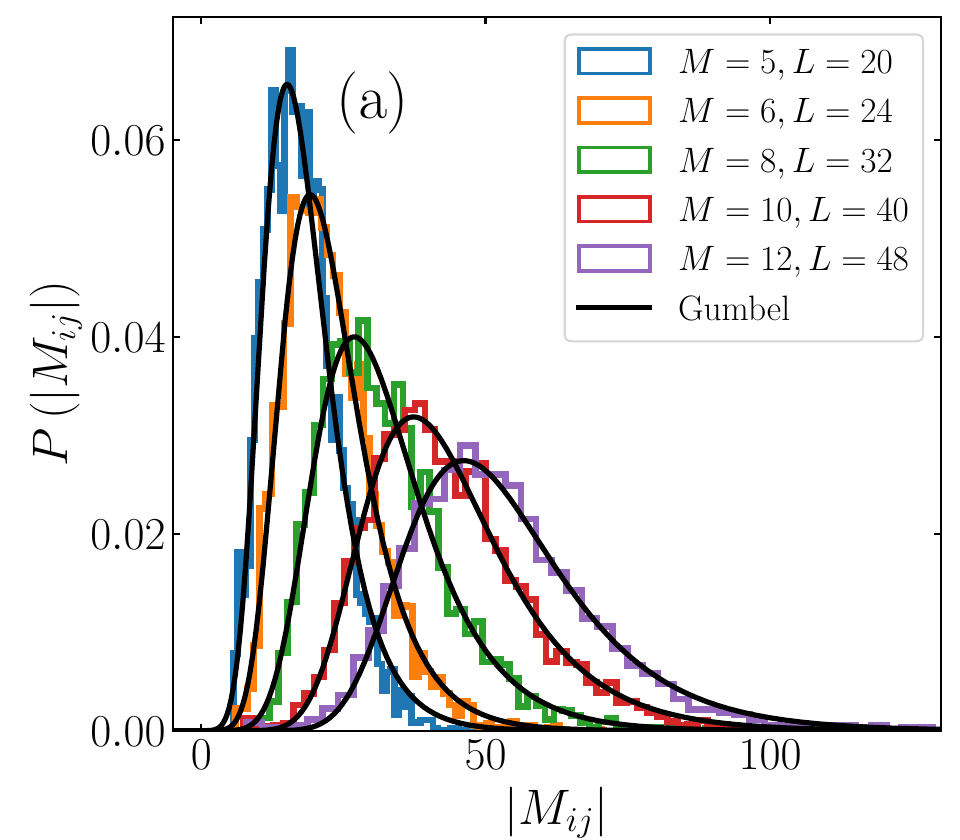}
    \includegraphics[width=.32\linewidth]{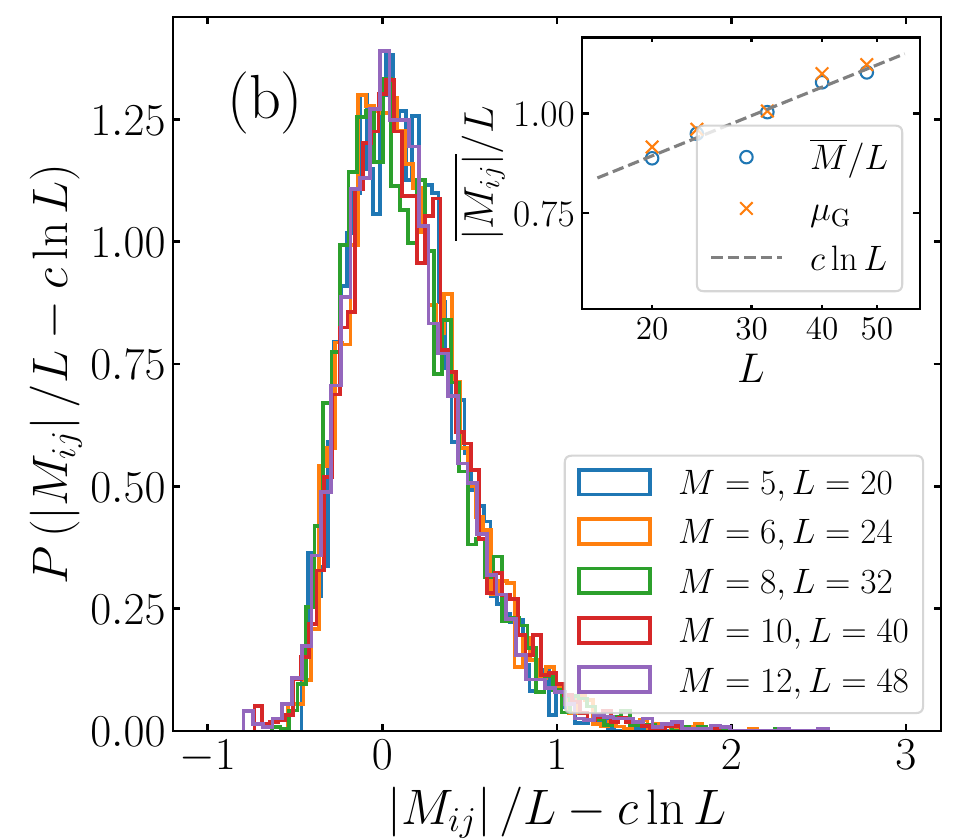}
    \includegraphics[width=.32\linewidth]{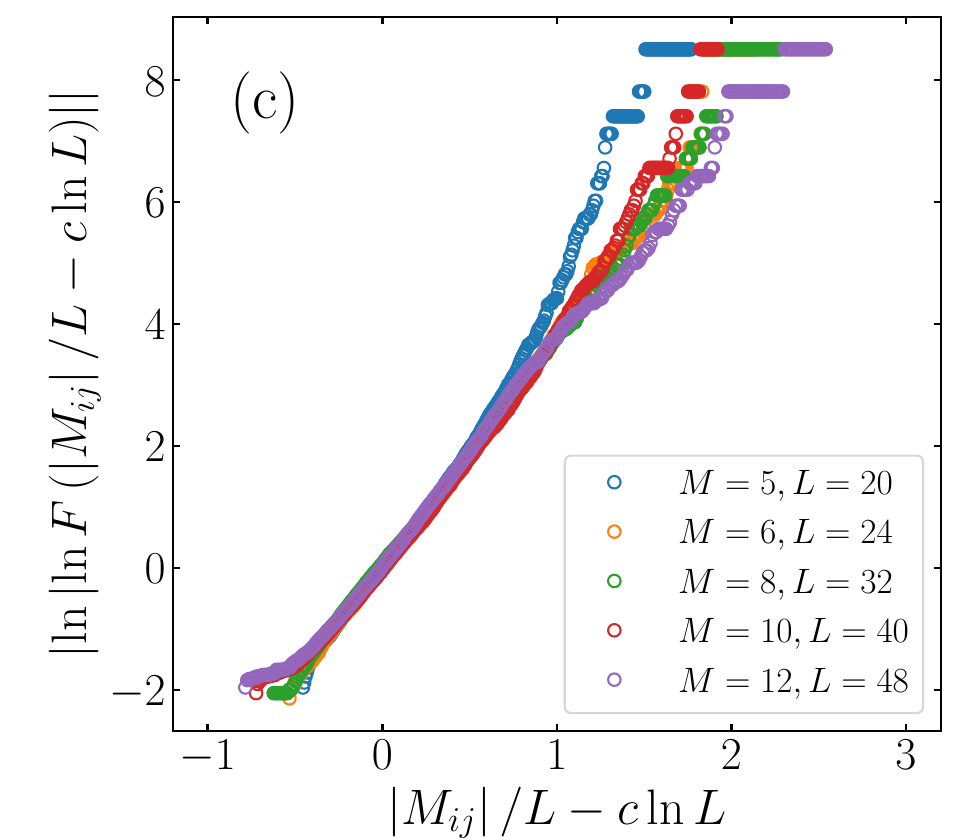}
    \caption{Distribution of $M_{ij}$ in \cref{eq:Mdef} of the two-spin 
    operator {$E_i^{(11)} E_{i+1}^{(22)}$} for $L = 20, 24, 32, 40,$ and $48$ and $M = L/4$.
    (a)~Histogram of the absolute value $|M_{ij}|$. The continuous lines are fits to the Gumbel 
    distribution~\eqref{eq:Gumbel}, with $\mu,\beta$ fitting parameters. 
    (b)~Histograms of $\abs{M_{ij}}/L- c\ln(L)$, with $c$ obtained by fitting the average 
    $\overline{M_{ij}}/L$ to the expected scaling~\eqref{eq:fit}. The fit is reported in the 
    inset with the gray-dashed line. We report with the cross symbols the average of the Gumbel 
    distributions $\mu_G=\mu+\beta\gamma$, with $\mu,\beta$ obtained from the fits in panel (a), and 
    $\gamma$ the Euler-Mascheroni constant. 
    (c)~Double logarithm of the \emph{CDF} of $\abs{M_{ij}}/L - c \ln(L)$. 
    }
    \label{fig:histogram_2spin_rat4_same_macro}
\end{figure*}

An important question is how universal is the statistics of off-diagonal matrix elements. 
To investigate this aspect, we consider the infinite-temperature thermal ensemble at fixed particle density $M/L=1/2$. Numerical data are reported in~\cref{fig:histogram_1spin_rat2_same_macro}. The qualitative behavior is similar to that at $M/L=1/4$ (see~\cref{fig:histogram_1spin_rat4_same_macro}). 
The continuous lines in~\cref{fig:histogram_1spin_rat2_same_macro}~(a) are fits to the Gumbel distribution~\eqref{eq:Gumbel}, and are in good 
agreement with the numerical data. 
In the inset of \cref{fig:histogram_1spin_rat2_same_macro}~(a) we report the rescaled histograms of 
$\abs{M_{ij}}/L$. 
The collapse is not perfect, similar to the case with $M/L=1/4$. 
We employ the same strategy as for $M/L=1/4$, i.e., fitting the average $\overline{M_{ij}}/L$ to~\eqref{eq:fit}, 
and plotting in Fig.~\ref{fig:histogram_1spin_rat2_same_macro} the shifted $M_{ij}/L-c\ln(L)$. 
The fit to the expected scaling~\eqref{eq:fit} gives $c \approx 0.16, c_0 \approx 0.94$ (see the inset 
in panel (b)). 
It is clear that the probability distribution $P({M}_{ij}-c\ln(L))$ is different 
than that for $M/L=1/4$. The data collapse for different $L$ is better than for $M/L=1/4$,  
and deviations near the peak become smaller upon increasing $L$. 
A fit of the histogram for $L=80$ to the Gumbel~\eqref{eq:Gumbel} gives $\beta\approx 0.29, \mu \approx 0.77$. 
Finally,  in~\cref{fig:histogram_1spin_rat2_same_macro}~(c) we  report the double logarithm of the numerical \emph{CDF} 
for $\abs{M_{ij}}/L - c \ln L$. As for $M=1/4$, in the central region the 
\emph{CDF} exhibits a clear linear behavior, as expected. 

\fed{Finally, we consider  the macrostate at inverse temperature $\beta = 0.5$, and particle density $M/L = 1/4$ in   \cref{fig:histogram_1spin_rat4_same_macro_beta0.5}.
Again we find that the qualitative behavior is similar to the infinite temperature case in \cref{fig:histogram_1spin_rat4_same_macro}.
In \cref{fig:histogram_1spin_rat4_same_macro_beta0.5}~(a) we fit the distribution of $|M_{ij}|/L$ to the Gumbel distribution~\eqref{eq:Gumbel}, observing a good agreement.
In the inset of \cref{fig:histogram_1spin_rat4_same_macro_beta0.5}~(a) we report the histogram of the rescaled matrix elements $\abs{M_{ij}}/L$.
As it was the case for the infinite temperature macrostate, rescaling with $L$ is not sufficient to obtain a perfect collapse. 
Moreover, as we show in the inset of \cref{fig:histogram_1spin_rat4_same_macro_beta0.5}~(b), the expected logarithmic behavior of the average $\overline{|M_{ij}|/L}$ cannot be easily extracted in the finite temperature case. 
For this reason, in the main plot of \cref{fig:histogram_1spin_rat4_same_macro_beta0.5} we shift the histograms by the averages $\overline{\abs{M_{ij}}}/L$, rather than by the fit as in \cref{fig:histogram_1spin_rat4_same_macro,fig:histogram_1spin_rat2_same_macro}.
We observe that after subtracting the average, the histograms present a perfect collapse for all values of $L$ except for $L = 160$. The discrepancy for $L=160$ is likely due to finite-size corrections. 
In \cref{fig:histogram_1spin_rat4_same_macro_beta0.5}~(c) we  report the double logarithm of the numerical \emph{CDF} of the subtracted $M_{ij}/L-\overline{\abs{M_{ij}}}/L$.
As for the infinite temperature macrostates, the \emph{CDF} exhibits a linear behavior, as expected from the Gumbel~\eqref{eq:GumbelCDF}.
}

\subsubsection{Two-spin operator}
\label{sec:two-spin}

Let us now discuss the statistics of off-diagonal matrix elements of local operators that have support on two nearest-neighbor sites. In the following we consider the infinite-temperature macrostate at $M/L=1/4$.
Moreover, since we employ the approach of Ref.~\cite{AlbaETHdiagonal}, as stressed already we are restricted to the macrostate constructed from eigenstates of the $XXX$ chain that do not contain bound states. 
Moreover, since the computational cost for evaluating the \emph{ABA} formulas (see \cref{app:matrixElements}) 
increases exponentially with the size of the operator support, 
we provide results for $L\le 60$. 
We anticipate that, despite this limitation, the qualitative scenario outlined for the one-spin operators remains 
the same. Let us focus on the operator {$E_{i}^{(11)}E_{i+1}^{(22)}=(\mathds{1}_i-\sigma^z_i)(\mathds{1}_{i+1}+\sigma^z_{i+1})/4$}, 
which is sensitive to anti-aligned spin configurations on nearest-neighbor sites. 

In \cref{fig:histogram_2spin_rat4_same_macro}~(a) we report the histogram of $M_{ij}$. 
The continuous lines are fits to the Gumbel distributions~\eqref{eq:Gumbel}, which are in 
perfect agreement with the numerical data. In Fig.~\ref{fig:histogram_2spin_rat4_same_macro}~(b) 
we show the histograms of the shifted $M_{ij}-c\ln(L)$, with $c$ obtained by fitting the 
average $\overline{M_{ij}}/L$ to~\eqref{eq:fit} (see the inset in panel (b)). The fit gives 
 $c \approx 0.25,c_0 \approx 0.15$.  
Despite the smaller chain sizes as compared with  \cref{fig:histogram_1spin_rat4_same_macro,fig:histogram_1spin_rat2_same_macro}, the data exhibit a 
reasonable collapse. A fit to the Gumbel distribution~\eqref{eq:Gumbel} gives $\beta \approx 0.28, \mu \approx 0$. 
Finally, in \cref{fig:histogram_2spin_rat4_same_macro}~(c) 
we plot the double logarithm of the \emph{CDF} of $\widetilde{M}_{ij}$.
Again, the approximate linear behavior in the central region 
confirms that $M_{ij}$ are well described by the Gumbel distribution. 

\subsubsection{Non-Gaussianity of off-diagonal matrix elements}
\label{sec:non-G}

Our results for off-diagonal matrix elements confirm that, despite integrability, the finite-size scaling 
is similar to the \emph{ETH} scaling~\eqref{eq:ETH}, even quantitatively. 
Precisely, the exponential decay as $e^{-L}$ is the same, except for the 
correction~\eqref{eq:fit}. The difference 
between integrable and non-integrable models emerges strikingly in the statistics 
of $M_{ij}$.  According to the ETH~\eqref{eq:ETH}, 
in chaotic models $R_{ij}$ follow a Gaussian  
distribution with zero mean and variance that depends on the energies $E_i$ and $E_j$ 
of the two eigenstates. On the other hand, we showed that in integrable systems \redd{the distribution  of $\widetilde{M}_{ij}$ is well described by a Gumbel distribution with parameters that are indipendent of $L$. This is not compatible with the assumption of a Gaussian distribution for the matrix elements, as we now discuss. 
}

To investigate the effects of the  non Gaussianity of \redd{the matrix elements},
following Ref.~\cite{leblond2019entanglement}, we compute the ratio $\Gamma$ defined as 
\begin{equation}\label{eq:gammaratio}
    \Gamma = \frac{\overline{\abs{\mathcal{O}_{ij}}^2}}{\overline{\abs{\mathcal{O}_{ij}}}^2}\,.
\end{equation}
Here $\mathcal{O}_{ij}$ is the matrix element between the eigenstates with energy $E_i$ and $E_j$, and $\overline{|\mathcal{O}_{ij}|^2}$ and $\overline{|\mathcal{O}_{ij}|}$ are  
the averages of $|\mathcal{O}_{ij}|^2$ and $|\mathcal{O}_{ij}|$ calculated over all 
the eigenstates with energies $E_i$ and $E_j$ such that $|E_i-E_j|\le \Delta E$. Here we 
choose the energy window $\Delta E=1.25$. 
For zero-mean Gaussian distributed $\mathcal{O}_{ij}$, the ratio $\Gamma$ is equal to $\pi/2$, 
and it is independent of the variance. This has been verified by using exact diagonalization 
in several nonintegrable systems (see, for instance, Ref.~\cite{leblond2019entanglement}).

\begin{figure}[t]
    \centering
    \includegraphics[width=.95\linewidth]{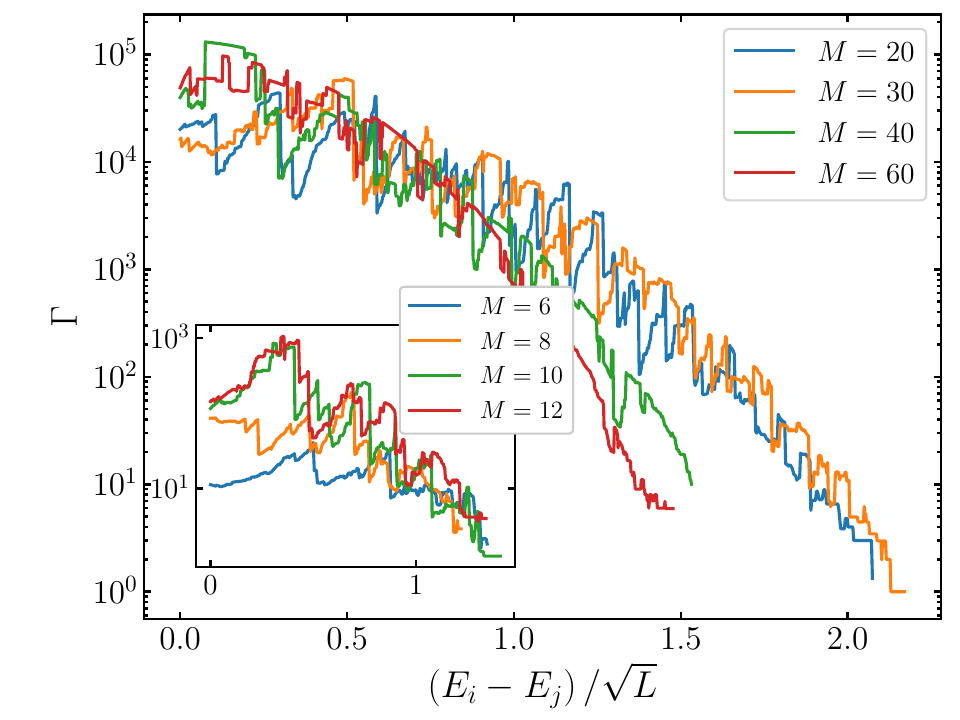}
    \caption{Value of the ratio $\Gamma$ in \cref{eq:gammaratio} for the operator $E_{i}^{(11)}$ for $M = 20, 30, 40,$ (cf.~\eqref{eq:elementary}) and $60$ and $L = 4M$ (main plot). In the inset we show the ratio $\Gamma$ for the two-spin operator $E_i^{(11)}E^{(22)}_{i+1}$.
    }
    \label{fig:ratios}
\end{figure}

In \cref{fig:ratios} we plot the ratio~\eqref{eq:gammaratio} as a function of the energy difference $(E_i - E_j)/\sqrt{L}$. 
In the main plot, we  consider the matrix elements $\mathcal{O}_{ij}$ of the 
one-spin operator $E_i^{(11)}$, whereas  in the inset those of the two-spin one $E^{(1,1)}_iE_{i+1}^{(2,2)}$ (cf.~\eqref{eq:elementary}). 
Clearly,  $\Gamma$ differs significantly from $\pi/2$, for both operators. Actually,  
we observe $\Gamma\sim 10^5$ in the region with $E_i\approx E_j$. 
We should mention that while the finite-size scaling of $\Gamma$ in the Heisenberg chain was already investigated in Ref.~\cite{leblond2019entanglement} by employing exact diagonalization, here we provide numerical data 
for much larger size $L\le 240$. 
The fact that $\Gamma\sim 10^5$ 
suggests that $\Gamma$ diverges in the thermodynamic limit $L\to\infty$. Indeed, we observe that 
$\Gamma$ grows upon increasing the total number of off-diagonal matrix elements employed to 
compute~\eqref{eq:gammaratio}. This is compatible with the fact that  
$\widetilde M_{ij}$ is well described by a Gumbel distribution, as we now show. First, we can 
neglect the logarithmic 
correction~\eqref{eq:fit} because it depends weakly on $|\mathcal{O}_{ij}|$ and it cancels out in the 
ratio~\eqref{eq:gammaratio}. A straightforward calculation allows us to obtain from~\eqref{eq:Gumbel} 
the distribution of $|O_{ij}|$ as 
\begin{equation}
\label{eq:Py}
P(y)=\frac{1}{\beta L  y}e^{-e^{\mu/\beta} y^{1/(\beta L)}}e^{\mu/\beta}y^{1/(\beta L)}.
\end{equation}
Now, one can compute $\Gamma$ in~\eqref{eq:gammaratio}. Clearly, one obtains that both 
the numerator and the denominator in~\eqref{eq:gammaratio} vanish in the limit $L\to\infty$. 
However, one can check that the ratio $\Gamma$ 
diverges exponentially in the limit $L\to\infty$, consistent with the observation in Fig.~\ref{fig:ratios}.

Conversely, it is interesting to derive the distribution of $-M_{ij}/L$ under the assumption that the off-diagonal matrix elements follow a Gaussian distribution. For instance, let us first consider the case in which they have real and imaginary parts that are uncorrelated and  Gaussian distributed with zero mean and variance $\sigma^2$. 
Now one  obtains that the distribution of $-M_{ij}/L$ is a  Gumbel $P_{\mu,\beta}$ with parameters 
\begin{equation}
\label{eq:G-fake}
\mu=-\frac{\ln(\sigma^2)}{L},\quad\beta=L^{-1}. 
\end{equation}
Since the variance vanishes exponentially with increasing $L$ (cf.~\eqref{eq:ETH}), $\mu$ becomes finite in the large $L$ limit. 
However, one has $\beta=1/L$, and, crucially, $\beta$ does not depend on $\sigma$. This implies that in the large $L$ limit the distribution~\eqref{eq:G-fake} becomes degenerate. This is in contrast with the results of Section~\ref{sec:one-spin}  and Section~\ref{sec:two-spin}, which provided numerical evidence that  the parameter $\beta$ is finite and nonzero in the large $L$ limit. 
It is intriguing to observe that this result  is in agreement with the general scenario  that the difference between integrable and nonintegrable models is in the size of the fluctuations of matrix elements, those being larger in integrable systems. Finally, if, instead, the matrix elements obey~\eqref{eq:ETH} with $R_{ij}$ real and Gaussian distributed, then it is straightforward to check that the distribution of $-M_{ij}/L$ is not of the Gumbel form. 

\subsection{Statistics of matrix elements between eigenstates involving different macrostates}
\label{sec:2-macro}

\begin{figure}    
    \centering
    \includegraphics[width=.95\linewidth]{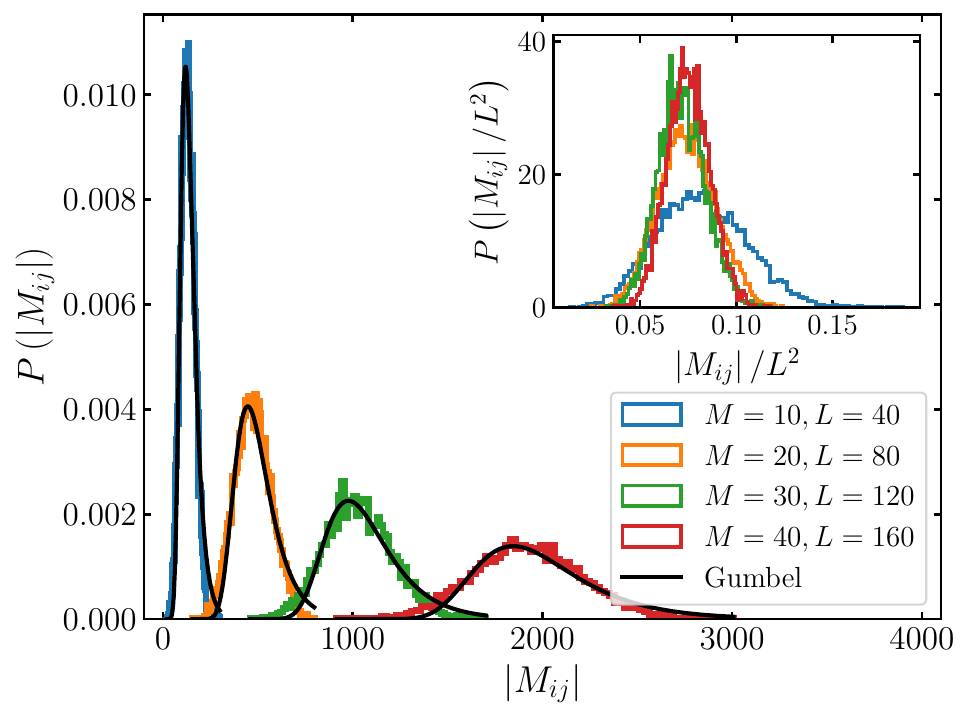}
    \caption{Distribution of the logarithm of the matrix element $|M_{ij}|$ of the one-spin operator $E_i^{(11)}$ between eigenstates in different macrostates. 
    In the inset we report the histogram of $|M_{ij}|$ rescaled with $L^2$. 
    The data for $L=120$ and $160$ collapse on the same curve.
    }
    \label{fig:histogram_different_macro1}
\end{figure}

\begin{figure}    
    \centering
    \includegraphics[width=.95\linewidth]{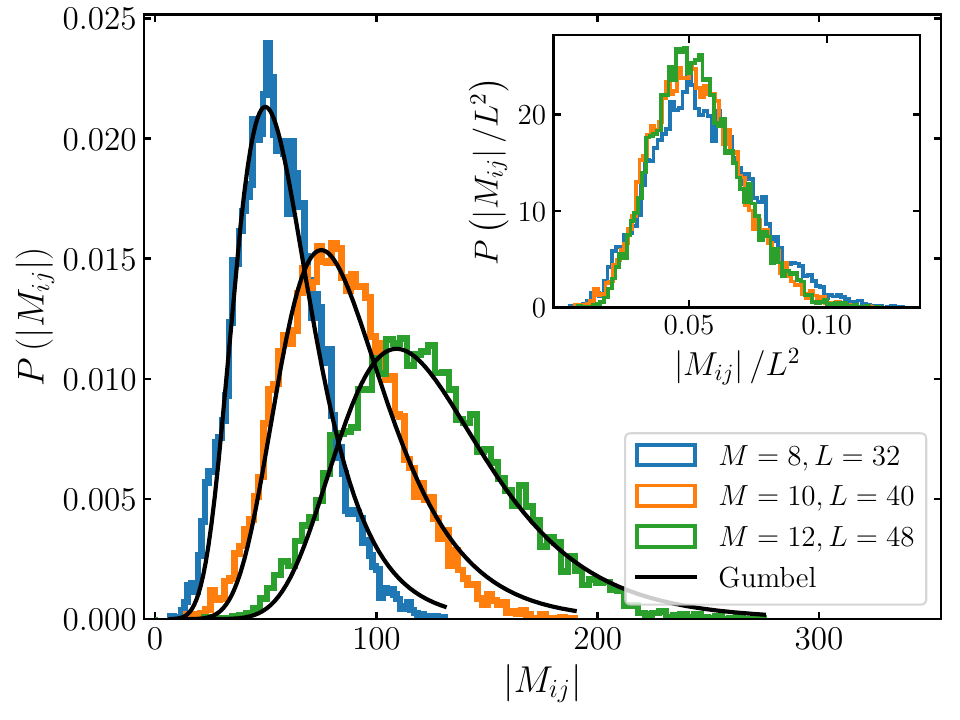}
    \caption{Distribution of the logarithm of the matrix element $|M_{ij}|$ of the two-spin operator {$E_i^{(11)} E_{i+1}^{(22)}$} between eigenstates in different macrostates, for $L = 32, 40,$ and $48$ and $M = L/4$.
    In the main plot we report the histogram of the absolute value of $\abs{M_{ij}}$. 
    In the inset we plot the logarithm $\abs{M_{ij}}$ rescaled by $L^2$, observing a collapse.
    }
    \label{fig:histogram_different_macro2}
\end{figure}

Let us now discuss off-diagonal matrix elements between eigenstates that are in different  thermodynamic macrostates.
\fed{Notice that the results that we present here go beyond the standard ETH framework, which applies to matrix elements between eigenstates in the same microcanonical window.}
First, exact results for hard-core bosons and numerical ones for the Lieb-Liniger gas (see~\cite{Essler:2023nhu}) suggest that off-diagonal matrix elements decay as 
\begin{equation}
\label{eq:asy-diff}
\langle E_i|\mathcal{O}|E_j\rangle \sim \exp\left( -{M_{ij}'}^{\scriptscriptstyle{\mathcal O}}L^2 \right),
\end{equation}
where  ${M_{ij}'}^{\scriptscriptstyle{\mathcal O}}$ depends on the observable 
and on the two macrostates. 
Here we focus on the infinite-temperature and the zero-temperature (i.e., the ground state) 
macrostates. Indeed, we numerically observe that it is in general challenging to reach the 
asymptotic regime $L\to\infty$, where~\eqref{eq:fab2} is expected to hold. This could be attributed to the  
 eigenstates-to-eigenstates fluctuations within the same macrostate at finite size $L$.  
These fluctuations give rise to  ``overlaps'' between the two macrostates, which make 
challenging to observe the asymptotic scaling~\eqref{eq:asy-diff}. It is natural to expect that the choice of the macrostates 
with $\beta=0$ and $\beta=\infty$  minimizes such finite-size corrections.
\fed{It is  important to stress that, while the ground state is represented by a legit macrostate in thermodynamic Bethe Ansatz (TBA) language, it is a zero entropy one.}

In  Fig.~\ref{fig:histogram_different_macro1} we show the histograms of the off-diagonal matrix elements  
$M_{ij}=\ln(|\langle E_i|{\mathcal O}|E_j\rangle|^2)$ for the one-spin operator  $E_i^{(11)}$ (cf.~\eqref{eq:elementary}).  
As in Section~\ref{sec:same}, we show results for $L = 40, 80, 120,$ and $160$ and fixed 
particle density $M = L/4$.  
By comparing with \cref{fig:histogram_1spin_rat4_same_macro}~(a) we observe that the typical value of 
$M_{ij}\approx 10^3$ is much larger than for off-diagonal matrix elements between eigenstates 
in the  same macrostate, which is compatible with the faster exponential decay~\cref{eq:fab2}. To extract the probability distribution 
function of $M_{ij}$, we fit the data to the Gumbel distribution~\eqref{eq:Gumbel}. The fits are reported 
with the continuous lines. The agreement between the data and the fits is not as good as for matrix elements 
between eigenstates in the same macrostate, although this could be a finite $L$ effect. 
To test the expected scaling~\eqref{eq:fab2}, in the inset of \cref{fig:histogram_different_macro1} we plot
$\abs{M_{ij}}/L^2$. The figure shows that all the data for $L\ge 80$ collapse on the 
same curve.  
In \cref{fig:histogram_different_macro2} we perform a similar analysis for the two-spin operator {$E_i^{(11)} E_{i+1}^{(22)}$}, showing data for  $L = 32, 40,$ and $48$ and $M = L/4$. 
Again, the qualitative behavior of the data is similar as in Fig.~\ref{fig:histogram_different_macro1}, 
in agreement with~\cref{eq:fab2}. 

\section{Conclusions}\label{sec:conclusions}

We investigated the \emph{ETH} scenario for off-diagonal matrix elements in the $\frac{1}{2}$-spin $XXX$ chain, which is 
a paradigmatic example of integrable lattice many-body quantum system. By employing state-of-the-art 
Algebraic Bethe Ansatz formulas, we studied off-diagonal matrix elements involving one-spin and two-spin operators  up to chain sizes $L\le 240$, which are far beyond state-of-the-art exact diagonalization results. We focused on the statistics of off-diagonal matrix elements between 
eigenstates that are both in the same thermodynamic macrostate, as well as eigenstates in different macrostates. 
Our results confirm the finite-size scaling observed in $1D$ integrable field theory in Ref.~\cite{Essler:2023nhu}. 
Precisely, off-diagonal matrix elements between eigenstates in the same macrostate decay exponentially with the 
chain size $L$, as in the standard \emph{ETH} scenario for chaotic models (cf.~\eqref{eq:ETH}). Off-diagonal 
matrix elements between eigenstates in different macrostates decay faster (cf.~\eqref{eq:fab2}). 
Interestingly, integrability is reflected in the statistics 
of the matrix elements, which is  encoded in the distribution of the $R_{ij}$ in~\eqref{eq:ETH}. 
In the standard \emph{ETH} scenario,  $R_{ij}$ follow a Gaussian distribution with zero mean. 
On the other hand,  we showed that for the $XXX$ chain 
the distribution is well described by a Gumbel distribution. 

Our work opens several new directions for future investigations. For instance, it would be interesting 
to employ our method to investigate the relationship between the statistics of off-diagonal matrix elements 
and Free Probability Theory (\emph{FPT}). The most urging question is  to what extent \emph{FPT} results~\cite{pappalardi2022eigenstate} also apply to integrable models.
Second, it was shown recently that the \emph{ETH} scenario, complemented with the \emph{FPT}, can be applied to describe the statistics of the overlaps between pre-quench initial states and 
eigenstates of chaotic systems~\cite{foini2025out}. It would be interesting to check this scenario. This is accessible by combining our method with the results of Ref.~\cite{alba2016the}. Finally, from the statistics of the matrix elements and the overlaps it is possible, in principle, to reconstruct the full-time dynamics of observables. 
This direction was investigated 
for the Lieb-Liniger model in Ref.~\cite{denardis2015relaxation}. It would be interesting to extend this results to lattice integrable systems. This would also allow to investigate 
relaxation dynamics and its relationship with \emph{ETH}~\cite{maceira2024thermalization,wang2025eigenstate,fritzsch2025micro,bouverotdupuis2025random,capizzi2025hydrodynamics}. 
\\
\section*{Acknowledgments}
We would like to thank Fabian Essler for useful discussions. We also thank the $GGI$ Florence 
for hospitality during the ``SFT 2025 - Lectures on Statistical Field Theories'', during which part of this work was done. 
This study was carried out within the National Centre on HPC, Big Data and Quantum Computing 
- SPOKE 10 (Quantum Computing) and received funding from the European Union Next-GenerationEU - 
National Recovery and Resilience Plan (NRRP) – MISSION 4 COMPONENT 2, INVESTMENT N. 1.4 – 
CUP N. I53C22000690001. This work has been supported by the project “Artificially 
devised many-body quantum dynamics in low dimensions - ManyQLowD” funded by the MIUR Progetti 
di Ricerca di Rilevante Interesse Nazionale (PRIN) Bando 2022 - grant 2022R35ZBF.

\bibliography{bibliography}

\newpage
\appendix

\section{Matrix elements of local operators in the XXZ chain}\label{app:matrixElements}

In this Appendix, we review the exact result of Refs.~\cite{Kitanine:1998oii,Kitanine:2000srg} for the matrix elements of the elementary operators~\eqref{eq:elementary} between energy eigenstates of the $XXZ$ spin-$1/2$ chain.
The Hamiltonian of the model is 
\begin{equation}\label{eq:XXZHam}
    H = \frac{1}{4} \sum_{i = 1}^{L} \frac{1}{2} \left ( \sigma_{i}^{+} \sigma_{i+1}^{-} + \sigma_{i}^{-} \sigma_{i+1}^{+} \right ) + \Delta\, \sigma_{i}^{z} \sigma_{i+1}^{z}\,,
\end{equation}
where, as in the main text, $\vec{\sigma}_i$ are the Pauli matrices and the chain has length $L$.
The parameter $\Delta$ is the anisotropy of the spin-spin interaction  and, in the isotropic limit 
$\Delta = 1$, the model~\eqref{eq:XXZHam} reduces to the $XXX$ spin chain~\eqref{eq:XXXHam} studied in the main text.
The Hamiltonian~\eqref{eq:XXZHam} has a global $U(1)$ symmetry corresponding to the conservation of the total magnetization $S^z = 1/2\sum_{i=1}^{L}  \sigma_i^z$. 
As mentioned in the main text, at $\Delta =1$ this symmetry is enhanced to the 
full $SU(2)$ group of global spin rotations.
In the following, in order to treat both the case $\Delta=1$ and $0<\Delta<1$ in an unified way, we define
\begin{align}
    \phi(\lambda) &= \begin{cases}
        \lambda, &\Delta = 1,\\
        \sinh{\lambda}, &0 < \Delta < 1,
    \end{cases}\\
    \eta &= \begin{cases}
        -\ii, &\Delta = 1,\\
        -\arccosh{\Delta}, &0 < \Delta < 1 .
    \end{cases}
\end{align}
To obtain  the energy spectrum of the Hamiltonian~\eqref{eq:XXZHam} and 
to compute matrix elements of operators we will use the \emph{ABA} formalism~\cite{korepin1993quantum}. 
In this formalism, for each site, one  introduces an $R$-matrix, which for the $XXZ$ chain 
takes the form
\begin{equation}\label{eq:Rmat}
    R_{0i}(\lambda-\xi) = \begin{pmatrix}
        1   &0  &0  &0\\
        0   &\frac{\phi(\lambda-\xi)}{\phi(\lambda-\xi+\eta)}   &0  &0\\
        0   &0  &\frac{\phi(\eta)}{\phi(\lambda-\xi+\eta)}  &0\\
        0   &0  &0  &1
    \end{pmatrix}.
\end{equation}
The $R$-matrix~\eqref{eq:Rmat} acts on the tensor product of the Hilbert space 
$\hilb_i$ of the $i$-th site and on an auxiliary space $\hilb_0$ isomorphic to $\CC^2$.
The so-called  monodromy matrix $T$ is then defined as the product in the auxiliary space 
of the $R$-matrices in each site as 
\begin{equation}\label{eq:monodromy}\begin{split}
    T(\lambda; \{\xi_i\}) &= R_{0N}(\lambda-\xi_N) \ldots R_{01}(\lambda-\xi_1)\\
    &= \begin{pmatrix}
        A(\lambda; \{\xi_i\})  &B(\lambda; \{\xi_i\})\\
        C(\lambda; \{\xi_i\})  &D(\lambda; \{\xi_i\})
    \end{pmatrix},
\end{split}\end{equation}
where the matrix $T$ acts on the product of the auxiliary $\hilb_0 \simeq \CC^2$ space and the Hilbert space $\hilb = \bigotimes_{i = 1}^L \hilb_i$ of the whole chain, while $A, B, C$ and $D$ are operators acting only on $\hilb$.
Following Refs.~\cite{Kitanine:1998oii,Kitanine:2000srg}, we have furthermore introduced arbitrary inhomogeneities $\{\xi_i\},\, i = 1, \ldots, L$ sitting at each site $i$.

The monodromy matrix~\eqref{eq:monodromy} allows us to construct the Fock space of the model.
Starting from the pseudovacuum state $\ket{0}$, which for the $XXZ$ chain is the ferromagnetic 
state with all spins up $\ket{0} = \ket{\uparrow\uparrow\uparrow\uparrow\ldots}$, the matrices $B(\mu)$ and $C(\mu)$ in \cref{eq:monodromy} act on the pseudovacuum $\ket{0}$ respectively as creation and annihilation operators 
of particles with rapidities $\mu_i$ 
\begin{equation}\label{eq:fock}
    \ket{\{\mu\}} = \prod_{k=1}^{M} B(\mu_k) \ket{0}, \quad \bra{\{\mu\}} = \bra{0} \prod_{k=1}^{M} C(\mu_k)\,.
\end{equation}
We remark that not all particle states~\eqref{eq:fock} are eigenstates of the Hamiltonian~\eqref{eq:XXZHam} and, as a consequence, they are not all orthogonal to each other,  forming an overcomplete basis. 
To find the eigenstates of~\cref{eq:XXZHam}, we define the transfer matrix $\transf$ 
as the trace over the internal space of the monodromy matrix~\eqref{eq:monodromy} $\transf(\lambda) = \Tr T(\lambda; \{ \xi_i\}) = A(\lambda; \{ \xi_i\}) + D(\lambda; \{ \xi_i\})$. 
The transfer matrix evaluated at two generic rapidities $\lambda,\mu$ commute with each other 
$\left [ \transf(\lambda), \transf(\mu)\right ] = 0$, meaning that they can be diagonalized simultaneously. 
Since $\lambda,\mu$ are arbitrary, this implies that we can diagonalize $\transf(\lambda)$. 

Moreover, in the homogeneous limit $ \forall i,\, \xi_i \to \eta/2$, the eigenstates of 
$\transf$ are also eigenstates of the $XXZ$ Hamiltonian~\eqref{eq:XXZHam}.
Applying the transfer matrix to the states~\eqref{eq:fock}, one obtains that the 
rapidities of the Hamiltonian eigenstates must satisfy a set of non-linear 
equations known as Bethe equations.  In the presence of the inhomogeneities $\xi_i$ the Bethe equations 
take the form 
\begin{equation}\label{eq:BetheEqApp}
    d(\lambda_i) \prod_{j\neq i} \frac{\phi(\lambda_j-\lambda_i-\eta)}{\phi(\lambda_j-\lambda_i+\eta)} = 1, 
\end{equation}
where the function $d(\lambda)$ is defined as
\begin{equation}\label{eq:ddefApp}
    d(\lambda) = \prod_{i = 1}^{L} \frac{\phi(\lambda-\xi_i)}{\phi(\lambda-\xi_i+\eta)}.
\end{equation}
Notice in particular that \cref{eq:ddefApp} vanishes when computed on the inhomogeneities, i.e., $d(\xi_i) = 0$.
The results for the $XXZ$ chain are obtained by taking the homogeneous limit $\xi_i \to \eta/2$. 
Importantly, the homogeneous limit is plagued by fictitious singularities that have to be 
removed to recover physical results. 

Moreover, as discussed already in Section~\ref{sec:model}, 
the Bethe equations~\eqref{eq:BetheEqApp} admit solutions containing both real and complex rapidities. 
Solving the Bethe equations for generic complex rapidities is a daunting task. 
However, according to the string hypothesis~\cite{takahashi1999thermodynamics}, in the thermodynamic limit 
$L \to \infty$, the vast majority of complex solutions organizes in groups of $n$ rapidities 
known as Bethe string, which have identical real part and imaginary parts equally spaced by $\eta$.
Physically, these strings states correspond to bound states of $n$ excitations moving collectively.
In practice, at finite $L$ the string hypothesis is valid only approximately, 
and the rapidities in the same string exhibit deviations $\delta$ which are exponentially small in $L$.
Assuming that the string hypothesis holds exactly even at finite (large) $L$ allows one to simplify considerably the solution of \cref{eq:BetheEqApp}. By taking the limit of zero string deviations $\delta\to0$ in~\eqref{eq:BetheEqApp}, one 
obtains the Bethe-Gaudin-Takahashi equations~\eqref{eq:bgt} for the real parts of the strings. 
Again, the limit $\delta\to0$ introduces additional fictitious singularities in the \emph{ABA} formulas for the 
matrix elements. 

\subsection{Scalar product of Bethe states}
\label{app:scalar}

A crucial ingredient in the \emph{ABA} formulas for the matrix elements of local operators is the so-called 
Slavnov formula for scalar products between Bethe states~\cite{korepin1993quantum}. 
Let $\ket{\{\lambda\}}$ be a Bethe state, with $M$ rapidities satisfying the Bethe 
equations~\eqref{eq:BetheEqApp}, and let $\ket{\{\mu\}}$ be a state of the form~\eqref{eq:fock} 
with $M$ arbitrary parameters $\mu$. Clearly, the state $\ket{\{\mu\}}$ for generic $\{\mu\}$ 
is not an eigenstate of the $XXZ$ chain. 
The two states are generically not orthogonal and their scalar 
product is given by the Slavnov formula~\cite{Slavnov:1989uvz,korepin1993quantum} 
\begin{equation}\label{eq:Slavnov}
    \braket{\{\mu\}|\{\lambda\}} = \frac{\det \mathbf{S}}{\prod_{i = 1}^{M} \prod_{j<i} \phi(\mu_i - \mu_j)\,\phi(\lambda_j - \lambda_i)}\,,
\end{equation}
where $\mathbf{S}$ is the Slavnov matrix~\cite{Slavnov:1989uvz,korepin1993quantum}
\begin{equation}\label{eq:SlavnovMat}\begin{split}
    \mathbf{S}_{ab} 
    &= - G_b^+ \left[ K_{ab}^+ - K_{ab}^- R_b \right ],
\end{split}\end{equation}
and we have introduced 
\begin{gather}
    K_{ab}^\pm = \frac{\phi(\eta)}{\phi(\lambda_a-\mu_b)\,\phi(\lambda_a-\mu_b \pm \eta)}\,,\label{eq:defKappa}\\
    G_b^\pm = \prod_{i=1}^{M} \phi(\lambda_i - \mu_b \pm \eta)\,,\label{eq:defG}\\
    R_b = d(\mu_b) \prod_{i=1}^{M} \frac{\phi(\lambda_i - \mu_b - \eta)}{\phi(\lambda_i - \mu_b + \eta)} = \frac{G_b^-}{G_b^+}\,d(\mu_b)\,. \label{eq:defR}
\end{gather}
If $\{\mu\}$ are also solutions of the Bethe equations, 
$\ket{\{\mu\}}$ is an eigenstate of the $XXZ$ chain, and we 
can use the Bethe equation~\eqref{eq:BetheEqApp} to replace $d(\mu_b) = \prod_{i\neq b} \phi(\mu_i - \mu_b + \eta)/\phi(\mu_i - \mu_b - \eta)$ in the Slavnov matrix~\eqref{eq:SlavnovMat}.
As a consequence now $R_b$ becomes
\begin{equation}
    R_b = - \frac{G_b^-}{G_b^+} \frac{\widetilde{F}_b^+}{\widetilde{F}_b^-}\,,
\end{equation}
where we have defined
\begin{equation}\label{eq:defFtilde}
    \widetilde{F}_b^\pm = \prod_{i=1}^{M} \phi(\mu_i - \mu_b \pm \eta)\,.
\end{equation}
While the previous expressions are always well defined for states with only real rapidities, 
for Bethe strings~\cref{eq:defFtilde} vanishes.
The rows of the Slavnov matrix~\eqref{eq:SlavnovMat} are then singular, although the determinant should remain finite.
For simplicity, let us assume that $\{\mu\}$ contain only a single $n$-string, and let us 
rearrange $\{\mu\}$ such that the first $n$ rapidities are the ones forming the $n$-string.
Following~\cite{caux2005computation},  we introduce string 
deviations $\delta$ for the first $n$ rapidities $\{\mu\}$. Then, we sum the first $n$ 
rows of the matrix~\eqref{eq:SlavnovMat}. This operation removes the singularities without 
changing the determinant, allowing us to take the limit $\delta\to0$. 
After these manipulations, the columns corresponding to the $n$ strings in the Slavnov 
matrix \cref{eq:SlavnovMat} are modified as 
\begin{equation}\label{eq:SlavnovMatString}
    \mathbf{S}'_{ab}= \begin{cases}
        - G_b^+ K_{ab}^+,   &b < n,\\
        \begin{aligned}
            - G_n^+ \Bigg [& K_{an}^+ - K_{a1}^{-} \prod_{i= 1}^{n} R_i \,+\\
            &- \sum_{j= 2}^{n} \left ( \partial K_{aj}^+ \prod_{i = j}^{n} R_i \right ) \Bigg],
        \end{aligned}&b = n,
    \end{cases}
\end{equation}
where $b=1,2,\dots, n$, and the derivative of the expression $K_{ab}^+$ is
\begin{equation}
    \partial K_{ab}^+ = -\frac{\phi(\eta)\, \phi(2\lambda-\eta)}{\phi(\lambda)^2\, \phi(\lambda-\eta)^2}\,. 
\end{equation}
The remaining columns of \cref{eq:SlavnovMat} are unchanged.
Notice that in the case $n = 1$, i.e., for a real rapidity, \cref{eq:SlavnovMatString} reduces to \cref{eq:SlavnovMat}, as expected. The procedure outlined above is straightforwardly generalized to 
the case in which $\{\mu\}$ contain an arbitrary number of strings~\cite{caux2005computation} because different strings can be treated independently. 

Let us now discuss the norm of the Bethe states. 
The norm is obtained from the Slavnov formula~\eqref{eq:Slavnov} 
by taking the limit $\{\mu\} \to \{\lambda\}$ in the matrix~\eqref{eq:SlavnovMat}. 
One obtains the so-called Gaudin formula~\cite{korepin1993quantum} 
\begin{equation}\label{eq:Gaudin}
    \braket{\{ \lambda \} | \{ \lambda \}} =  \frac{\det \mathbf{G}}{\prod_{i = 1}^M \prod_{j\neq i} \phi(\lambda_i - \lambda_j)}\,,
\end{equation}
where the elements of the Gaudin matrix $\mathbf{G}$ are~\cite{korepin1993quantum}
\begin{equation}\label{eq:GaudinMat}\begin{split}
    \mathbf{G}_{ab} 
    =&\, -G_{b}^+\Bigg [ k^{(2)}(\lambda_a - \lambda_b)\, +\\
    &+\delta_{ab}\Bigg ( \frac{d'(\lambda_b)}{d(\lambda_b)} - \sum_{j = 1}^{M} k^{(2)}(\lambda_j - \lambda_b)\Bigg ) \Bigg ],
\end{split}\end{equation}
and we have introduced the notation 
\begin{equation}
    k^{(2)}(\lambda) = \frac{\phi(2\eta)}{\phi(\lambda+\eta)\, \phi(\lambda-\eta)}\,.
\end{equation}

Similar to the Slavnov matrix, the Gaudin matrix is singular if strings are present. 
The strategy employed to remove the singularities in the Slavnov matrix has 
to be modified because the limit $\{\mu\} \to \{\lambda\}$ and the limit $\delta \to  0$ do not commute. 
We introduce $F_b^\pm$ to be the finite part of $\widetilde{F}_b^\pm$ obtained by isolating the singular part in the limit $\delta\to 0$
\begin{align}
    F_b^+ &\approx \begin{cases}
        (\delta_b -\delta_{b+1})^{-1}\, \widetilde{F}_b^+, &b<n,\\
        \widetilde{F}_b^+, &b=n.
    \end{cases}  \\
    F_b^- &\approx \begin{cases}
        (\delta_{b-1} -\delta_{b})^{-1}\, \widetilde{F}_b^-, &1 < b\leq n,\\
        \widetilde{F}_b^-, &b=1.
    \end{cases}  
\end{align}
where $\delta_b$ is the deviation of the rapidity $\lambda_b$.
We obtain that the columns in the Gaudin matrix that correspond to a $n$-string are modified as 
\begin{align}
    &\mathbf{G}'_{ab} = \begin{cases}
        F_b^+,  &a = b,\\
        -F_b^+, &a = b+1,\\
        0,      &\text{otherwise,}
    \end{cases}\hspace{2cm} b < n, \notag  \\
    &\mathbf{G}'_{an} = \begin{cases}
        \begin{aligned}
            - F_n^+ \Bigg ( & \frac{d'(\lambda_a)}{d(\lambda_a)}\, +\\
            &- \sum_{j > n} k^{(2)}(\lambda_a - \lambda_j) \Bigg ),
        \end{aligned}  &a \leq n,\\
        - F_n^+ \sum_{j \leq n} k^{(2)}(\lambda_a - \lambda_j), &a > n,
    \end{cases}\label{eq:GaudinMatString}
\end{align}
where the sum over $j>n$ in $\mathbf{G}'_{an}$ is carried over all the rapidities that do not belong 
to the string, including possibly all other strings.
Again we see that for $n = 1$, \cref{eq:GaudinMatString} reduces to \cref{eq:GaudinMat}.

\subsection{Matrix elements of local operators}
\label{app:matrix}

We now provide the formulas for matrix elements of local operators between 
eigenstates of the $XXZ$ chain. In Section~\ref{app:gen-spin} we provide 
the general formula for operators having nontrivial support on $\ell$ 
consecutive sites. In Section~\ref{app:one-spin} we discuss the case with 
$\ell=1$, providing the result also for eigenstates that contain Bethe strings. 

\subsubsection{General formulas for $\ell$-spin operators}
\label{app:gen-spin}

In the main text we studied operators constructed from elementary ones $E_j^{(\alpha,\beta)}$ 
acting on a generic site $i$ of the chain as 
\begin{equation}\label{eq:elementaryApp}\begin{aligned}
    &E_i^{(11)} = \frac{1}{2}(\mathds{1}_i+\sigma^z_i),
    \quad
    &&E_i^{(12)} = 
  \frac{1}{2}(\sigma^x_i+i\sigma^y_i),\\
    &E_i^{(21)} = 
    \frac{1}{2}(\sigma_i^x-i\sigma_i^y),
    &&E_i^{(22)} 
    = \frac{1}{2}\left ( \mathds{1}_i-\sigma^z_i \right ).
\end{aligned} 
\end{equation}
In Refs.~\cite{Kitanine:1998oii,Kitanine:2000srg}, the authors demonstrated that the 
local operators~\eqref{eq:elementaryApp} can be written in terms of the monodromy 
matrix $T$~\eqref{eq:monodromy} and the transfer matrix $\transf = A + D$ as
\begin{equation}\label{eq:KMT}
    E_i^{(\epsilon_i, \epsilon'_i)} = \left (\prod_{j < i} \transf(\xi_j) \right ) T_{\epsilon_i, \epsilon'_i}(\xi_i) \left ( \prod_{j > i} \transf(\xi_j) \right ).
\end{equation} 
One can exploit the fact that the Bethe state $\ket{\{\lambda\}}$ is an eigenstate of the transfer matrix $\transf$. Indeed, by using \cref{eq:KMT} and by applying the transfer matrix to the right eigenstate, we can write
\begin{equation}
    \bra{\{\mu\}} \prod_{i=1}^\ell E_i^{(\epsilon_i, \epsilon'_i)} \ket{\{ \lambda\}} = \Phi_\ell(\{\lambda\})\,\bra{\{\mu\}} \prod_{i=1}^\ell T_{\epsilon_i,\epsilon_i'} \ket{\{\lambda\}},
\end{equation}
where
\begin{equation}
    \Phi_\ell(\{\lambda\}) = \prod_{j = 1}^\ell \prod_{a = 1}^M \frac{\phi(\lambda_a - \xi_j)}{\phi(\lambda_a - \xi_j + \eta)}\,,
\end{equation}
is the product of the eigenvalues of $\transf(\xi_j)$. 
To compute the matrix element we then apply the operators $A, B, C$ and $D$ 
in \cref{eq:monodromy} on the state $\bra{\{\mu\}}$ on the left.
Recalling the definition~\eqref{eq:fock} for the state $\bra{\{\mu\}}$, we observe 
that $C(\mu_{M+1})$ acts as a creation operator for  an additional excitation with 
rapidity $\mu_{M+1}$. However, the action of $A, B$ and $D$ is more 
complicated and it is given as~\cite{Kitanine:1998oii,Kitanine:2000srg,Alba:2009th}
\begin{widetext}
\begin{gather}
   \bra{0} \prod_{k = 1}^{M} C(\mu_k)\, A(\mu_{M+1})
   = \sum_{a' = 1}^{M+1} \frac{\prod_{k=1}^{M}\phi(\mu_k-\mu_{a'}+\eta)}{\prod_{k=1,\, k\neq a'}^{M+1} \phi(\mu_k-\mu_{a'})} \left ( \bra{0} \prod_{k = 1,\, k\neq a'}^{M+1} C(\mu_k) \right ),\label{eq:A}\\
   \bra{0} \prod_{k = 1}^{M} C(\mu_k)\, D(\mu_{M+1})= \sum_{a = 1}^{M+1} d(\mu_a) \frac{\prod_{k=1}^{M}\phi(\mu_k-\mu_{a}-\eta)}{\prod_{k=1,\, k\neq a}^{M+1} \phi(\mu_k-\mu_{a})} \left ( \bra{0} \prod_{k = 1,\, k\neq a}^{M+1} C(\mu_k) \right ),\label{eq:D}\\
\begin{split}
    \bra{0} \prod_{k = 1}^{M} C(\mu_k)\, B(\mu_{M+1})
    =& \sum_{a=1}^{M} d(\mu_a) \frac{\prod_{k=1}^{M}\phi(\mu_k-\mu_a-\eta)}{\prod_{k=1,\, k\neq a}^{M+1}\phi(\mu_k -\mu_a)} \times\\
    &\times\sum_{a'=1,\, a'\neq a}^{M+1} \frac{\prod_{k=1,\,k\neq a}^{M}\phi(\mu_k-\mu_{a'}-\eta)}{\prod_{k=1,\, k\neq a,\,a'}^{M+1}\phi(\mu_k -\mu_{a'})}\left ( \bra{0} \prod_{k = 1,\, k\neq a,\,a'}^{M+1} C(\mu_k) \right )\label{eq:B}.
\end{split}
\end{gather}
\end{widetext}
In \cref{eq:B}, we used the fact that in the case of interest $\mu_{M+1} = \xi$ and $d(\xi) = 0$ to simplify the expression~\cite{Kitanine:1998oii,Kitanine:2000srg,Alba:2009th}.
By recursively applying the outlined expressions, we obtain the action of a string of local 
operators on any Bethe state. 
To express the result in a closed form, we define the two set of indices $\alpha^\pm$ as~\cite{Kitanine:1998oii,Kitanine:2000srg,Alba:2009th}
\begin{equation}\begin{split}
    \alpha^+ &= \left \{ j : 1 \leq j \leq \ell,\, \epsilon_j  = + \right \},\\
    \alpha^- &= \left \{ j : 1 \leq j \leq \ell,\, \epsilon'_j = - \right \}.
\end{split}\end{equation}
For every index $j \in \alpha^+$ ($j \in \alpha^-$), we furthermore introduce a set $a_j$ ($a_j'$) such that
\begin{equation}\begin{split}
    1 \leq a_j \leq M+j,\, a_j \in \mathbf{A}_j\, \text{ for } j \in \alpha^-,\\
    1 \leq a'_j \leq M+j,\, a'_j \in \mathbf{A}'_j\, \text{ for } j \in \alpha^+,
\end{split}\end{equation}
where the sets $\mathbf{A}_j$ and $\mathbf{A}_j'$ are recursively defined as
\begin{equation}\begin{split}
    \mathbf{A}_j &= \big \{ b \in [1, M+\ell]: b \neq a_k,\, a_k' \text{ with } k < j \big \},\\
    \mathbf{A}_j' &= \big \{ b \in [1, M+\ell]: b \neq a_k \text{ with } k < j,\\
    &\hspace{3.2cm} \text{ and } b \neq a_k' \text{ with } k \leq j \big \}.
\end{split}\end{equation}
With these definitions, we can finally write 
\begin{equation}\begin{aligned}
    &\bra{0} \prod_{k = 1}^{M} C(\mu_k)\, \prod_{j=1}^{\ell} T_{\epsilon_j \epsilon_{j'}}(\mu_{M+j}) \\
    =& \sum_{\{a_j, a_j' \}} G_{\{a_j, a_j'\}}(\mu) \left ( \bra{0}\prod_{b \in \mathbf{A}_{s+1}} C(\mu_b) \right ),
\end{aligned}\end{equation}
where we defined
\begin{equation}\begin{aligned}
    G_{\{a_j, a_j'\}} = 
    \prod_{j \in \alpha^-}\! d(\mu_{a_j})\frac{\prod_{b=1,\,b\in \mathbf{A}_j}^{M+j-1} \phi(\mu_{b}-\mu_{a_j}-\eta)}{\prod_{b=1,\,b\in \mathbf{A}'_j}^{M+j}\phi(\mu_{b}-\mu_{a_j})} \times\\
    \times
    \prod_{j \in \alpha^+}\! \frac{\prod_{b=1,\,b\in \mathbf{A}'_j}^{M+j-1} \phi(\mu_{b}-\mu_{a'_j}+\eta)}{\prod_{b=1,\,b\in \mathbf{A}_{j+1}}^{M+j}\phi(\mu_{b}-\mu_{a'_j})} ,
\end{aligned}\end{equation}
in terms of the sets $\mathbf{A}_j$ and $\mathbf{A}_j'$. 

Finally in order to obtain the matrix elements we take the homogeneous limit $\xi_j \to \eta/2$.
As explained in Ref.~\cite{Alba:2009th}, however, this introduces 
singularities due to the terms that contain the operators $A$ and $B$. 
Moreover,  for most of the matrix elements, besides multiplicative singularities, 
there remain additional poles which cancel out between the different terms in the sum~\cite{Alba:2009th}. 
For states which only include real rapidities, the procedure to remove the singularities was discussed in 
Ref.~\cite{Alba:2009th}. However, in the presence of strings, the limit of 
vanishing string deviations and the inhomogeneous limit do not commute, and an efficient procedure to 
obtain the matrix elements between generic eigenstates of the $XXZ$ chain also in the presence of strings is 
not available yet. A notable exception is the case of one-spin operator, i.e., $\ell=1$, which we discuss in 
Section~\ref{app:one-spin}. 

\subsubsection{One-spin operators}
\label{app:one-spin}

For operators acting on a single spin the inhomogeneous limit does not introduce any singularity.  
Thus, the limit $\delta\to0$, i.e., of vanishing string deviations can be performed efficiently. 
Moreover, by using properties of rank-$1$ matrices, the matrix element of one-spin operators 
can be expressed as a single determinant, instead of a sum of multiple determinants. 
Precisely, we can use that if $P$ is a rank-$1$ matrix and $M$ is a generic one, 
the determinant of their sum is
\begin{equation}\label{eq:rank1}
    \det (M+P) = \det M + \sum_{j = 1}^{M} \det M^{(j)},
\end{equation}
where $M^{(j)}$ is the matrix obtained by replacing the $j$-th column of $M$ with the $j$-th column of $P$.

From \cref{eq:A} we observe that the matrix elements of $E_{i}^{(11)}$, which 
are proportional to  the matrix elements of $A$, are written as 
sum of determinants as in~\eqref{eq:rank1}. Each term in the sum is the determinant of a 
matrix that is obtained from the Slavnov matrix~\eqref{eq:Slavnov} (or the Gaudin matrix~\eqref{eq:GaudinMat} for diagonal matrix elements) by replacing a single column with a column of the matrix $P$ defined as~\cite{Kitanine:2000srg}
\begin{equation}\begin{split}
    \mathbf{P}_{ab} 
    &= \frac{\phi(\eta)\, \prod_{i = 1}^{M}\phi(\mu_i - \mu_b + \eta)}{\phi\!\left ( \lambda_a - \frac{\eta}{2} \right )\phi\!\left ( \lambda_a + \frac{\eta}{2} \right )}\\
    &= \frac{\phi(\eta)\, F^+_{b}}{\phi\!\left ( \lambda_a - \frac{\eta}{2} \right )\phi\!\left ( \lambda_a + \frac{\eta}{2} \right )}.
\end{split}\end{equation}
By using \cref{eq:rank1}, one obtains that the off-diagonal matrix elements of one-spin operators 
can then be written as
\begin{multline}
\label{eq:bra-ket}
    \bra{\{\mu\}} E^{(11)} \ket{\{\lambda\}} 
    =\, \frac{\prod_{i = 1}^{M} \phi\!\left ( \lambda_i + \frac{\eta}{2}\right )}{\prod_{i = 1}^{M} \phi\!\left ( \mu_i + \frac{\eta}{2}\right )} \\\times\frac{\det (\mathbf{S} + \mathbf{P})}{\prod_{i=1}^M\prod_{j<i} \phi(\lambda_i -\lambda_j)\, \phi(\mu_j - \mu_i)}, 
\end{multline}
where $\mathbf{S}$ is the Slavnov matrix. 
The diagonal matrix elements can be obtained by simply substituting the Slavnov matrix $\mathbf{S}$ with the 
Gaudin matrix $\mathbf{G}$. 
By using this result and the relation $E^{(22)} = \mathds{1} - E^{(11)}$, we also obtain 
$\bra{\{\mu\}} E^{(11)} \ket{\{\lambda\}} = \braket{\{\mu\}|\{\lambda\}} - \bra{\{\mu\}} E^{(22)} \ket{\{\lambda\}}$.

Having the matrix elements in the form~\eqref{eq:bra-ket}, it is now possible to 
treat eigenstates containing strings. 
Indeed, in the presence of an $n$-string $\{\mu_{1}, \ldots,\mu_{n}\}$, the 
Slavnov matrix $\mathbf{S}$ can be modified as in \cref{eq:SlavnovMatString}, by 
recursively adding the columns of the block corresponding to the string.
In order to preserve the determinant, the same operations have to be 
performed on the matrix $\mathbf{P}$, yielding 
\begin{equation}
    \mathbf{P}_{ab}' = \begin{cases}
        0\,,  &b < n,\\
        \begin{aligned}
        &\frac{\phi(\eta)\, G^+_{n}}{\phi\!\left ( \lambda_a - \frac{\eta}{2} \right )\phi\!\left ( \lambda_a + \frac{\eta}{2} \right )} \times\\
        &\hspace{1cm}\times\sum_{j=1}^{n} \frac{F^+_j}{G^+_j} \left ( \prod_{i=j+1}^{M} R_i \right ),
        \end{aligned} &b = n.
    \end{cases}
\end{equation}
Let us now discuss the matrix elements of $E_{i}^{(21)}$. Unlike $E_{i}^{(11)}$, $E_{i}^{(21)}$ connects 
eigenstates with different number of particles. Now, one can check that $E_i^{(21)}$ is  proportional to 
the creation operator $C(\eta/2)$. This means that the matrix elements between 
states $\ket{\{\mu_k\}}, j = 1, \ldots, M$ and $\ket{\{\lambda_j\}}, j = 1, \ldots, M+1$ is 
simply given by the formula
\begin{multline}
    \bra{\{\mu\}} E^{(12)} \ket{\{\lambda\}} =
    \frac{\Phi(\{\lambda\})}{\prod_{k=1}^{M}\phi(\mu_k-\eta/2)}\\
   \times \frac{\det \mathbf{S'}}{\prod\limits_{1\leq i < j \leq M+1} \phi(\lambda_i-\lambda_j) \prod\limits_{1\leq l<m \leq M} \phi(\mu_l-\mu_m)}
\end{multline}
where $\mathbf{S'}$ coincides with the Slavnov matrix $\mathbf{S}$ except for the $M+1$-th column, which is 
given as 
\begin{equation}
    \mathbf{S'}_{ab} = \begin{cases}
        \mathbf{S}_{ab}\,,  &b\leq M,\\
        -\phi(\eta)\, \frac{\prod_{j = 1}^{M+1}\phi(\lambda_j+\eta/2)}{\phi(\lambda_a-\eta/2)\, \phi(\lambda_a+\eta/2)}\,,  &b =M+1.
    \end{cases}
\end{equation}
The matrix elements of $E^{(21)}$ are then simply obtained using $E^{(21)} = E^{(12)\,\dagger}$.

\end{document}